\documentclass[12pt]{article}
\usepackage{epsfig}

\def\beq{\begin{equation}}
\def\eeq#1{\label{#1}\end{equation}}
\def\eeqn{\end{equation}}

\def\beqa{\begin{eqnarray}}
\def\eeqa#1{\label{#1}\end{eqnarray}}
\def\eeqan{\end{eqnarray}}

\let\bar=\overbar

\def\Dslash{\not{\hbox{\kern-4pt $D$}}}
\def\dslash{\not{\hbox{\kern-2pt $\del$}}}

\def\msb{{\bar{\ssstyle M \kern -1pt S}}}

\def\Title#1{\begin{center} {\Large {\bf #1} } \end{center}}

\begin{document}

\Title{Higgs boson search at ATLAS}

\bigskip\bigskip


\begin{raggedright}  

{\it Scott Snyder on behalf of the ATLAS Collaboration\index{Snyder, S.}\\
Brookhaven National Laboratory\\
Upton, NY, 11973, USA}
\bigskip\bigskip
\end{raggedright}

\section{Introduction}

The 2011 run of the Large Hadron Collider (LHC) at CERN was very successful,
with experiments accumulating nearly $5~{\mbox{fb$^{-1}$}}$ of data.  Of particular
interest was the search for a Standard Model (SM) Higgs boson, the remaining
component of the SM that has not yet been definitively
observed.  This note summarizes the status of this search at the
ATLAS experiment~\cite{Aad:2008zzm} using
the full 2011 data set of $(4.6$--$4.9)~{\mbox{fb$^{-1}$}}$.

\section{Higgs boson search channels}

The Higgs boson search is carried out for many different final states.
At the LHC, the dominant production mode for a Higgs boson
is gluon-gluon fusion ($gg\rightarrow H$).  For a
Higgs boson with mass ${\ensuremath{m_H}}$ above about $135{\ifmmode {\mathrm{\ Ge\kern -0.1em V}}\else\textrm{Ge\kern -0.1em V}\fi}$, the dominant
decay is into a pair of $W$ or $Z$~gauge bosons;
such decays that include leptons in the final state are distinctive
enough to search for directly.  For lower Higgs boson masses, the
dominant decay is into two $b$-quarks.  The very large heavy-flavour
multijet background makes it infeasible to search for this decay
directly.  One can look for rarer but more distinctive decays
of the Higgs boson, such as $H\rightarrow\gamma\gamma$
or $H\rightarrow\tau\tau$, for diboson decays in which one of the
gauge bosons is off-shell, or for a Higgs boson produced in association
with a $W$ or $Z$~boson.  The best sensitivity for ${\ensuremath{m_H}}\approx125{\ifmmode {\mathrm{\ Ge\kern -0.1em V}}\else\textrm{Ge\kern -0.1em V}\fi}$
is obtained from $H\rightarrow\gamma\gamma$, followed by
$H\rightarrow ZZ^{(*)}\rightarrow \ell\ell\ell'\ell'$.

\subsection{$H{\rightarrow}\gamma\gamma$ channel}

The diphoton decay of the Higgs~boson is relatively rare
(with a branching ratio $\sim 0.2\%$).  In order to have good
sensitivity in this channel~\cite{ATLAS:2012ad},
one must have very good $m_{\gamma\gamma}$
resolution as well as good control over non-photon backgrounds.
ATLAS achieves a mass resolution of about $1.7\%$ at ${\ensuremath{m_H}}=120{\ifmmode {\mathrm{\ Ge\kern -0.1em V}}\else\textrm{Ge\kern -0.1em V}\fi}$,
and the fraction of non-photon background is less than 30\%.

\begin{figure}[htb]
\begin{center}
{{%
  \newdimen\figsize
  \setlength\figsize{\hsize}%
  \addtolength\figsize{-2\columnsep}%
  \divide\figsize by 2
  \vbox{%
  \makebox{\parbox[t]{\figsize}{\vskip 0.1pt \epsfig{file=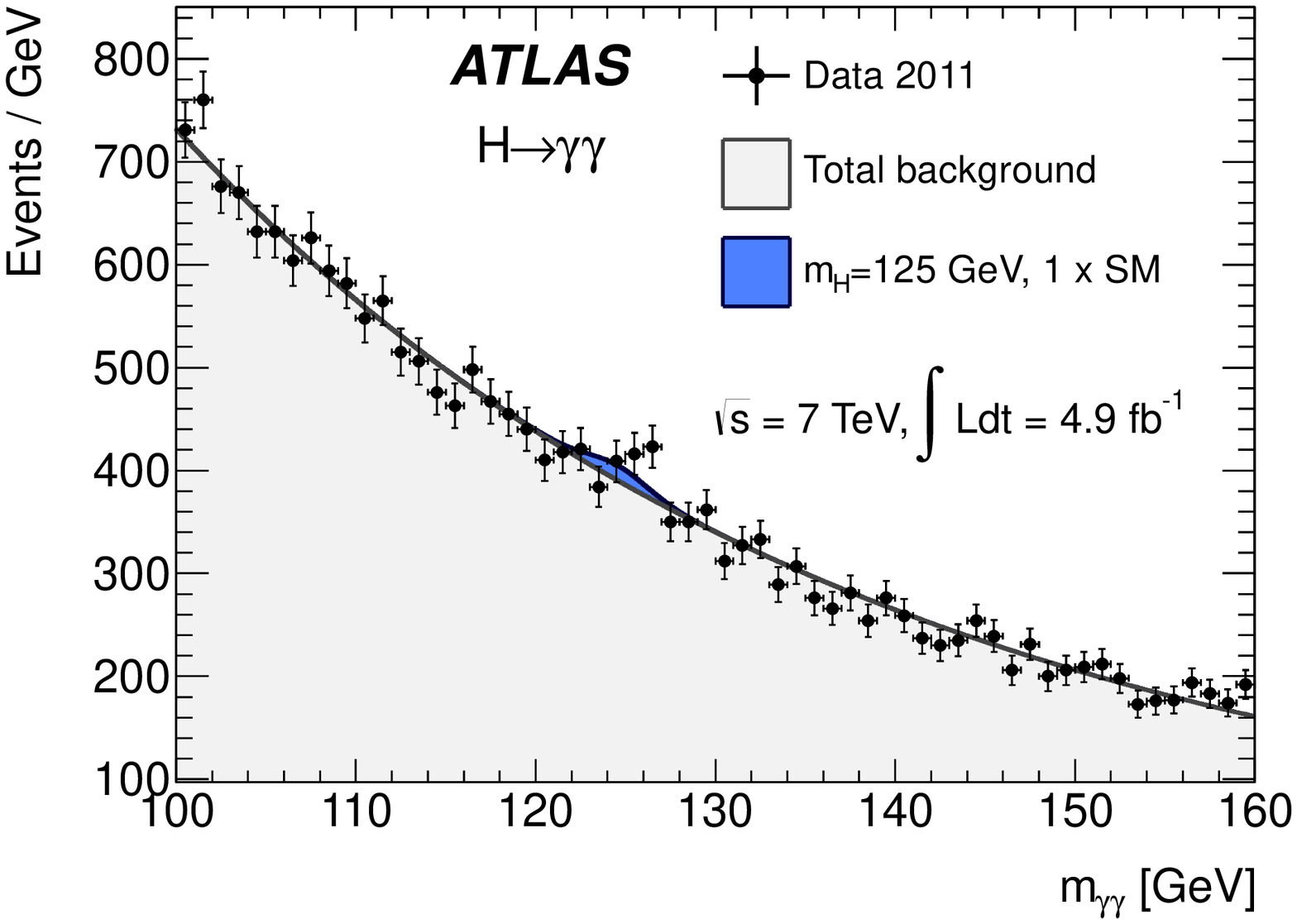, width=\figsize, height=3.6cm}}%
           \hspace{2\columnsep}%
           \parbox[t]{\figsize}{\vskip 0.1pt \epsfig{file=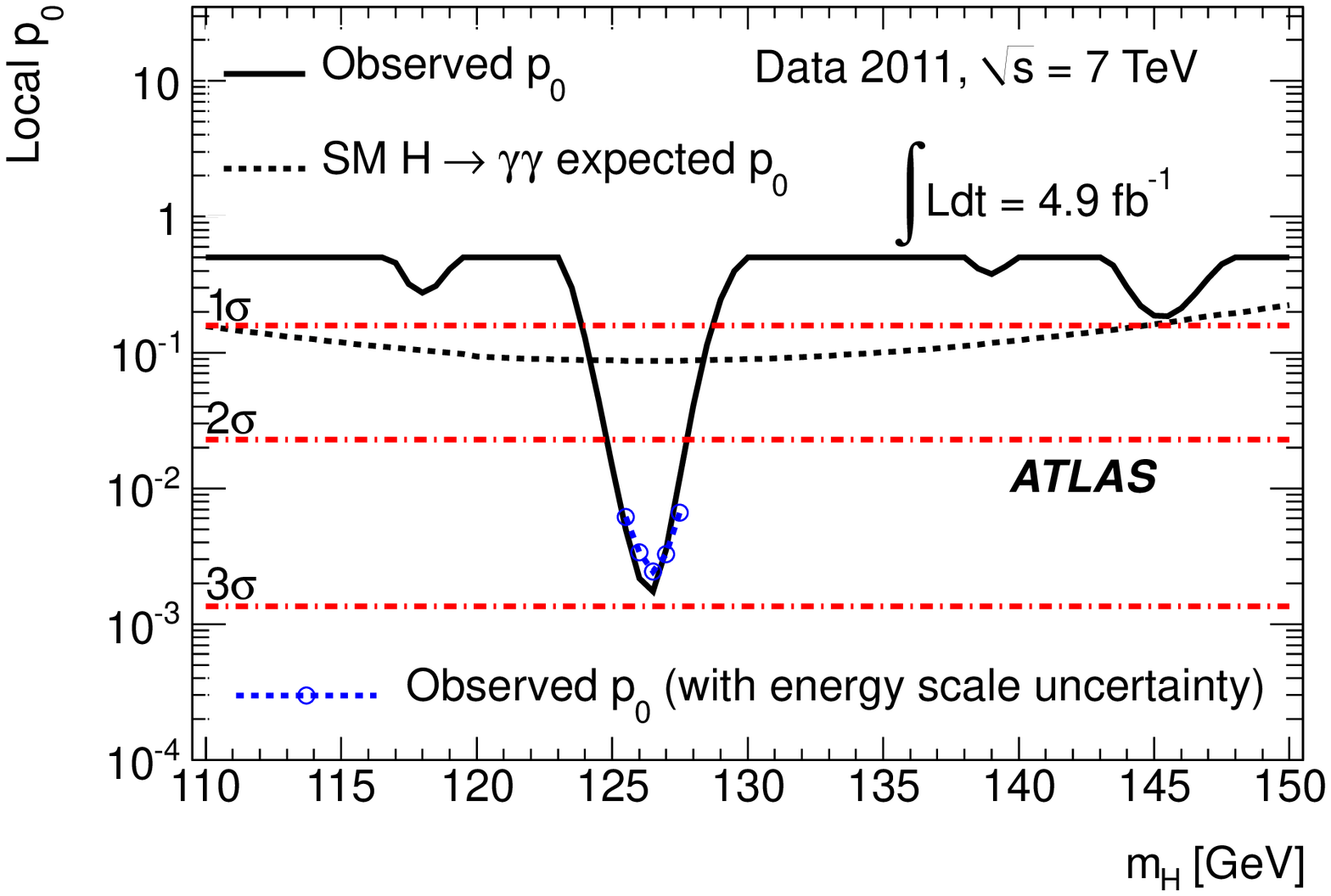, width=\figsize, height=3.6cm}}}}}}
\caption{Results from the $H{\rightarrow}\gamma\gamma$ channel~\cite{ATLAS:2012ad}.
  Left: Invariant
  mass distribution for the entire sample, overlaid with the fitted
  total background.  The expectation for a ${\ensuremath{m_H}}=125{\ifmmode {\mathrm{\ Ge\kern -0.1em V}}\else\textrm{Ge\kern -0.1em V}\fi}$ SM
  Higgs boson is also shown.  Right: Local probability $p_0$ for the background
  to fluctuate to the observed number of events or higher.  The dashed
  line shows the expected median local $p_0$ for the signal hypothesis
  when tested at ${\ensuremath{m_H}}$.}
\label{fig:gg}
\end{center}
\end{figure}

To optimize the expected sensitivity, the analysis is split into
nine subchannels with differing mass resolution and expected signal
fraction, depending on the kinematics of the photons and whether
they converted.  The background is modeled with an exponential fit
to the data.
Results are shown in Figure~\ref{fig:gg}.
The ${\ensuremath{m_H}}$ regions
$(113$--$115){\ifmmode {\mathrm{\ Ge\kern -0.1em V}}\else\textrm{Ge\kern -0.1em V}\fi}$ and $(134.5$--$136){\ifmmode {\mathrm{\ Ge\kern -0.1em V}}\else\textrm{Ge\kern -0.1em V}\fi}$ are excluded
at 95\% CL.  A small excess is seen
around ${\ensuremath{m_H}}=126.5{\ifmmode {\mathrm{\ Ge\kern -0.1em V}}\else\textrm{Ge\kern -0.1em V}\fi}$, with a local significance of $2.8\sigma$, or
a global significance of $1.5\sigma$ when the
look-elsewhere effect is taken into account over
the ${\ensuremath{m_H}}$ range $(110$--$150){\ifmmode {\mathrm{\ Ge\kern -0.1em V}}\else\textrm{Ge\kern -0.1em V}\fi}$.

\subsection{$H{\rightarrow} ZZ^{(*)}{\rightarrow} \ell\ell\ell'\ell'$ channel}

In this channel~\cite{ATLAS:2012ac},
one looks for two pairs of opposite-sign, same-flavour
leptons ($e$ or $\mu$), with one pair having an invariant mass
close to the $Z$~boson mass.  This channel has a quite small background
and a fully-reconstructed Higgs~boson decay, which allows it to have
good sensitivity over a wide range of Higgs~boson masses,
from $600{\ifmmode {\mathrm{\ Ge\kern -0.1em V}}\else\textrm{Ge\kern -0.1em V}\fi}$ down to $110{\ifmmode {\mathrm{\ Ge\kern -0.1em V}}\else\textrm{Ge\kern -0.1em V}\fi}$.  The background is primarily
$ZZ$~diboson production, with smaller contributions from $Z+{\mathrm{jets}}$
and ${t\bar t}$.

\begin{figure}[htbp]
\begin{center}
{{%
  \newdimen\figsize
  \setlength\figsize{\hsize}%
  \addtolength\figsize{-\columnsep}%
  \addtolength\figsize{-\columnsep}%
  \divide\figsize by 3
  \vbox{%
  \makebox{\parbox[t]{\figsize}{\vskip 0.1pt \epsfig{file=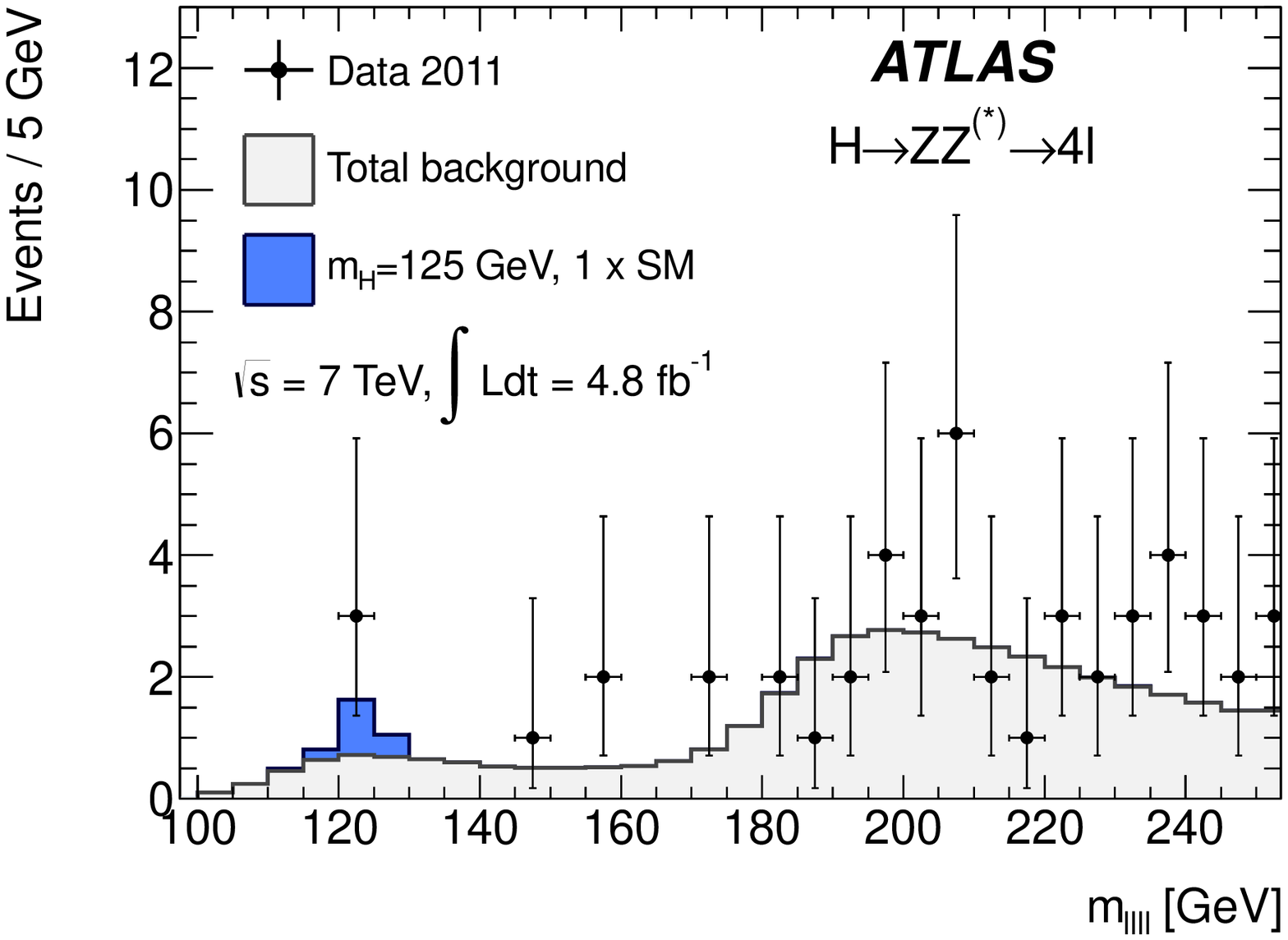, width=\figsize,height=3.8cm}}%
           \hspace{\columnsep}%
           \parbox[t]{\figsize}{\vskip 0.1pt \epsfig{file=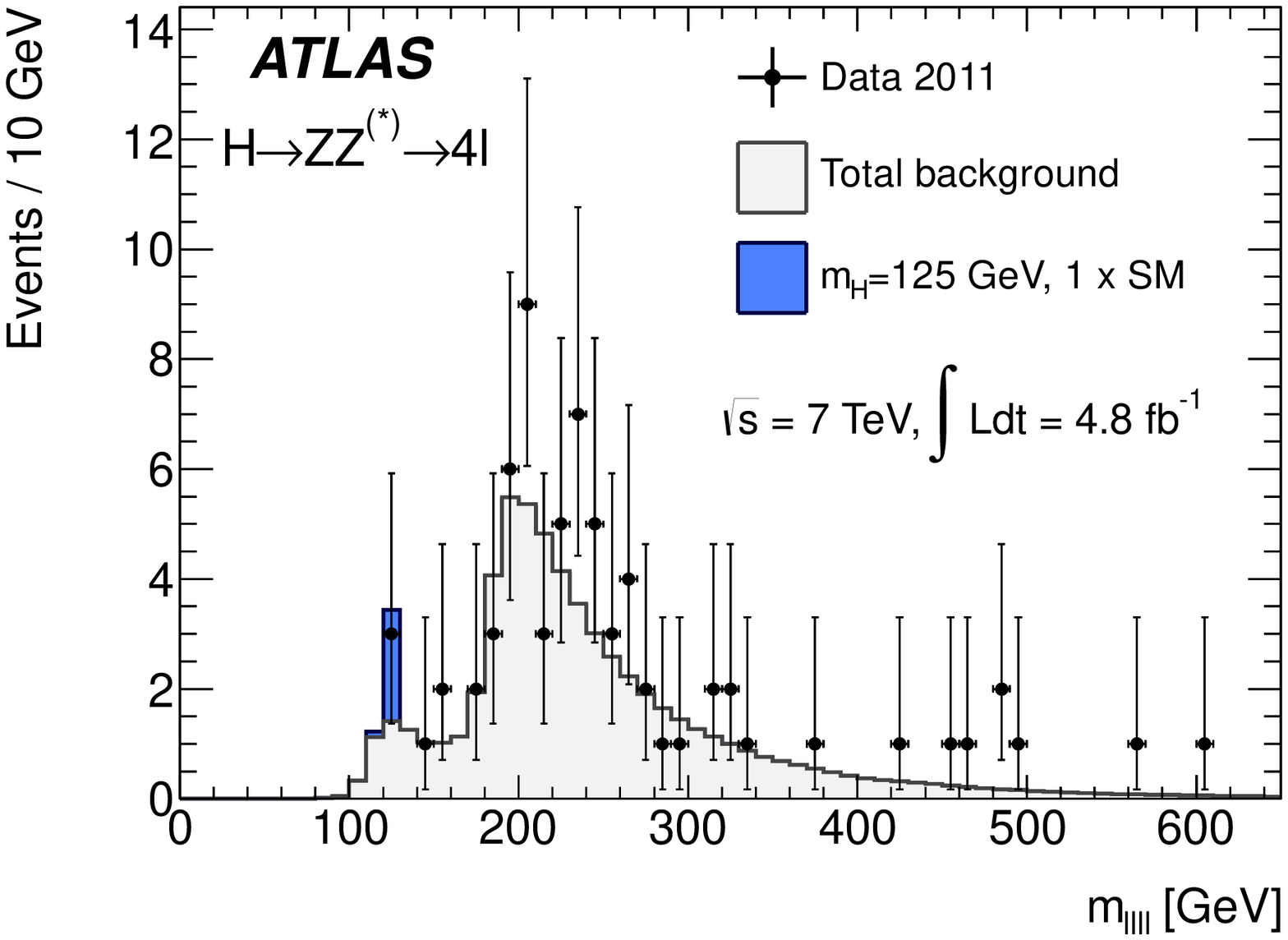, width=\figsize,height=3.8cm}}%
           \hspace{\columnsep}%
           \parbox[t]{\figsize}{\vskip 0.1pt \epsfig{file=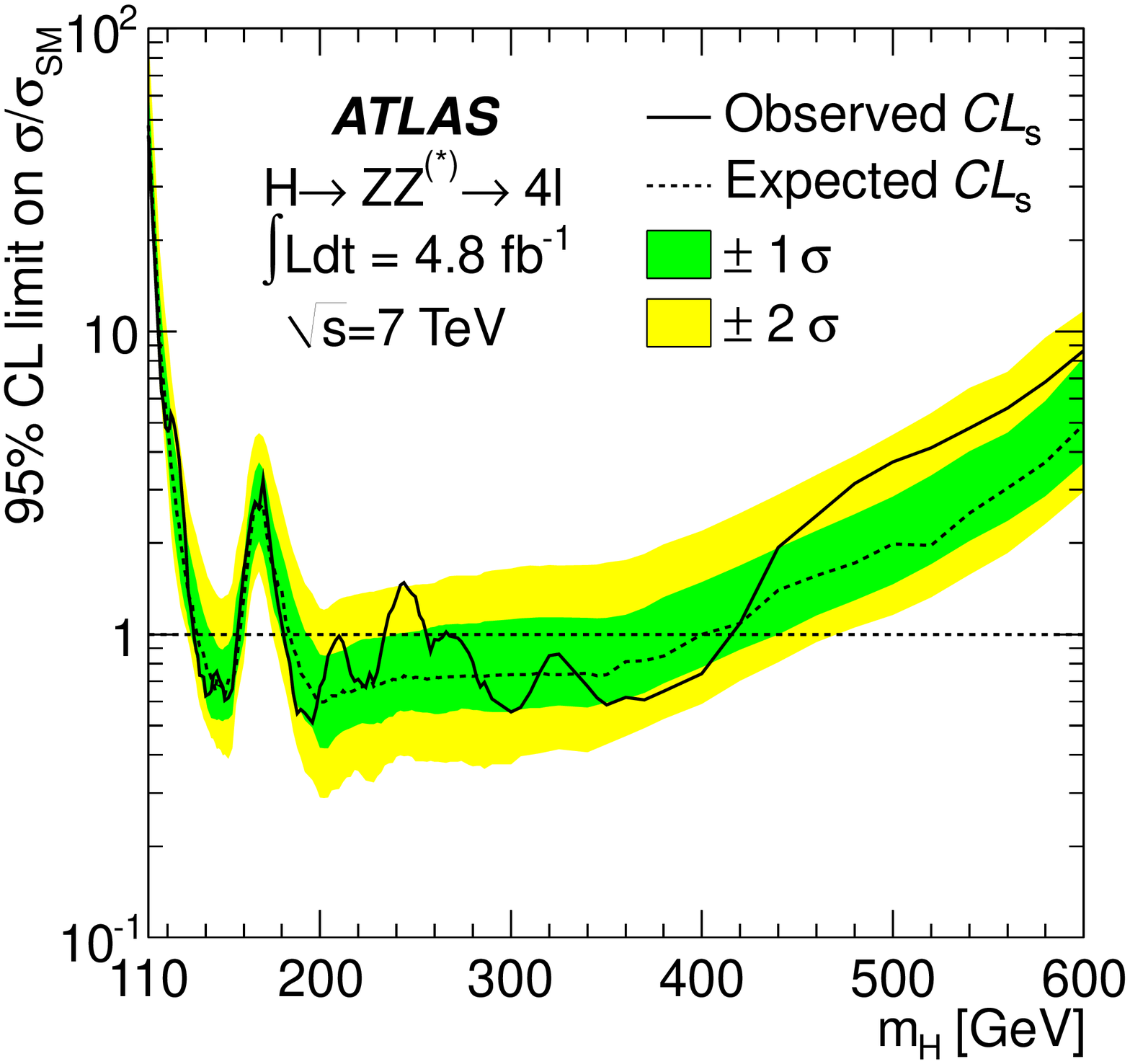, width=\figsize,height=3.8cm}}}}}}%
{{%
  \newdimen\figsize
  \setlength\figsize{\hsize}%
  \addtolength\figsize{-\columnsep}%
  \addtolength\figsize{-\columnsep}%
  \divide\figsize by 3
  \vbox{%
  \makebox{\parbox[t]{\figsize}{\vskip 0.1pt \epsfig{file=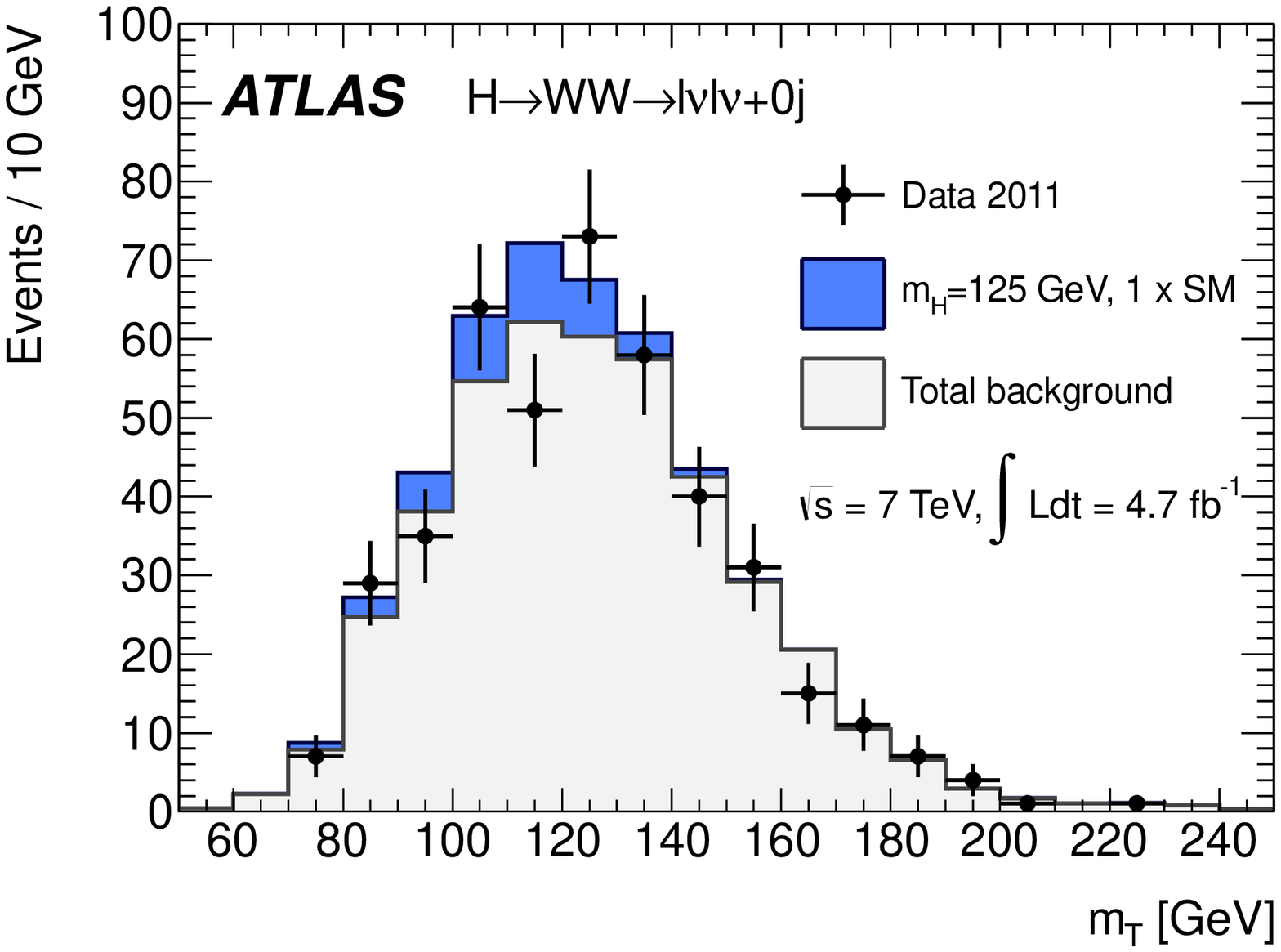,width=\figsize}}%
           \hspace{\columnsep}%
           \parbox[t]{\figsize}{\vskip 0.1pt \epsfig{file=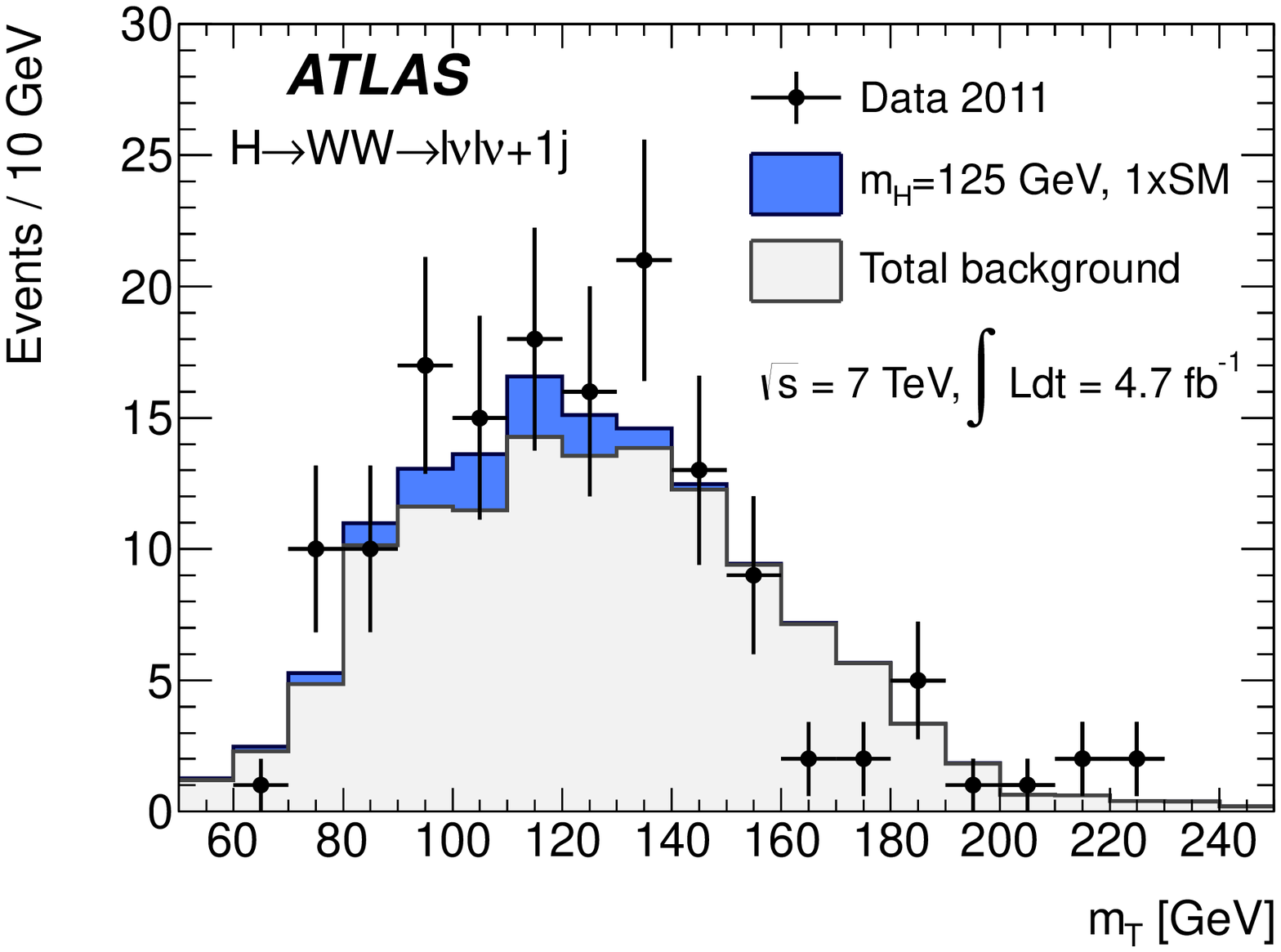,width=\figsize}}%
           \hspace{\columnsep}%
           \parbox[t]{\figsize}{\vskip 0.1pt \epsfig{file=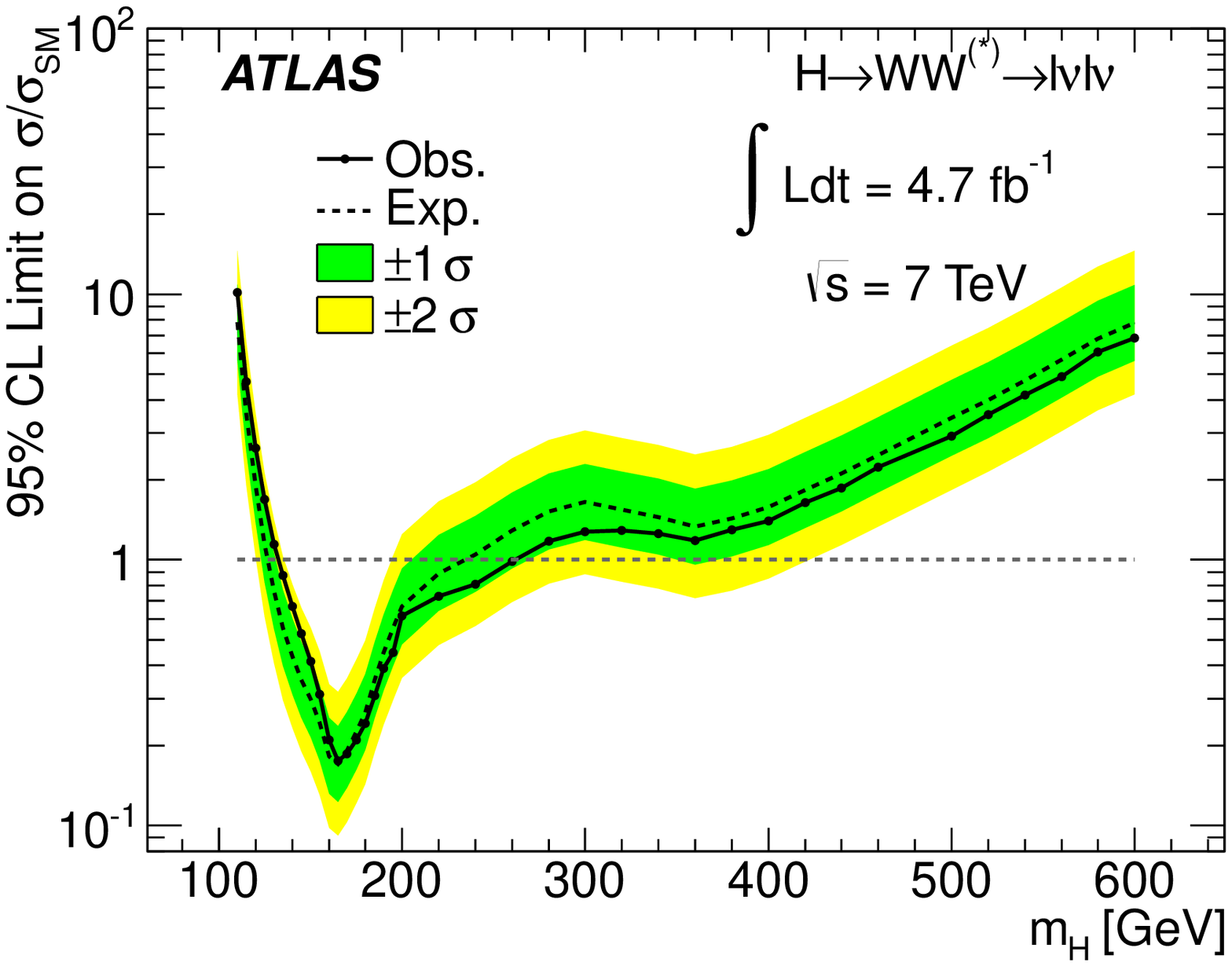,width=\figsize}}}}}}%
{{%
  \newdimen\figsize
  \setlength\figsize{\hsize}%
  \addtolength\figsize{-\columnsep}%
  \addtolength\figsize{-\columnsep}%
  \divide\figsize by 3
  \vbox{%
  \makebox{\parbox[t]{\figsize}{\vskip 0.1pt \epsfig{file=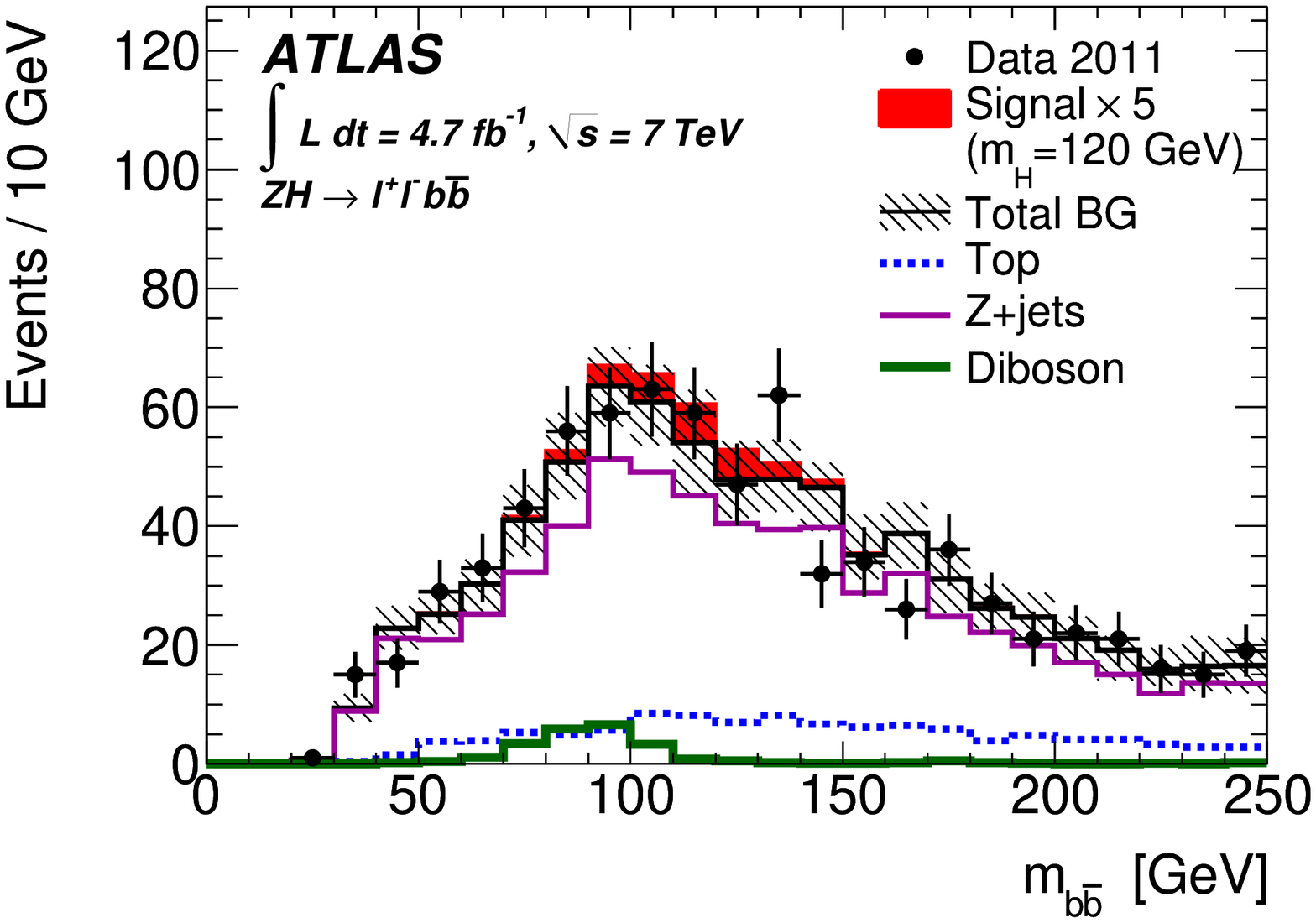,width=\figsize}}%
           \hspace{\columnsep}%
           \parbox[t]{\figsize}{\vskip 0.1pt \epsfig{file=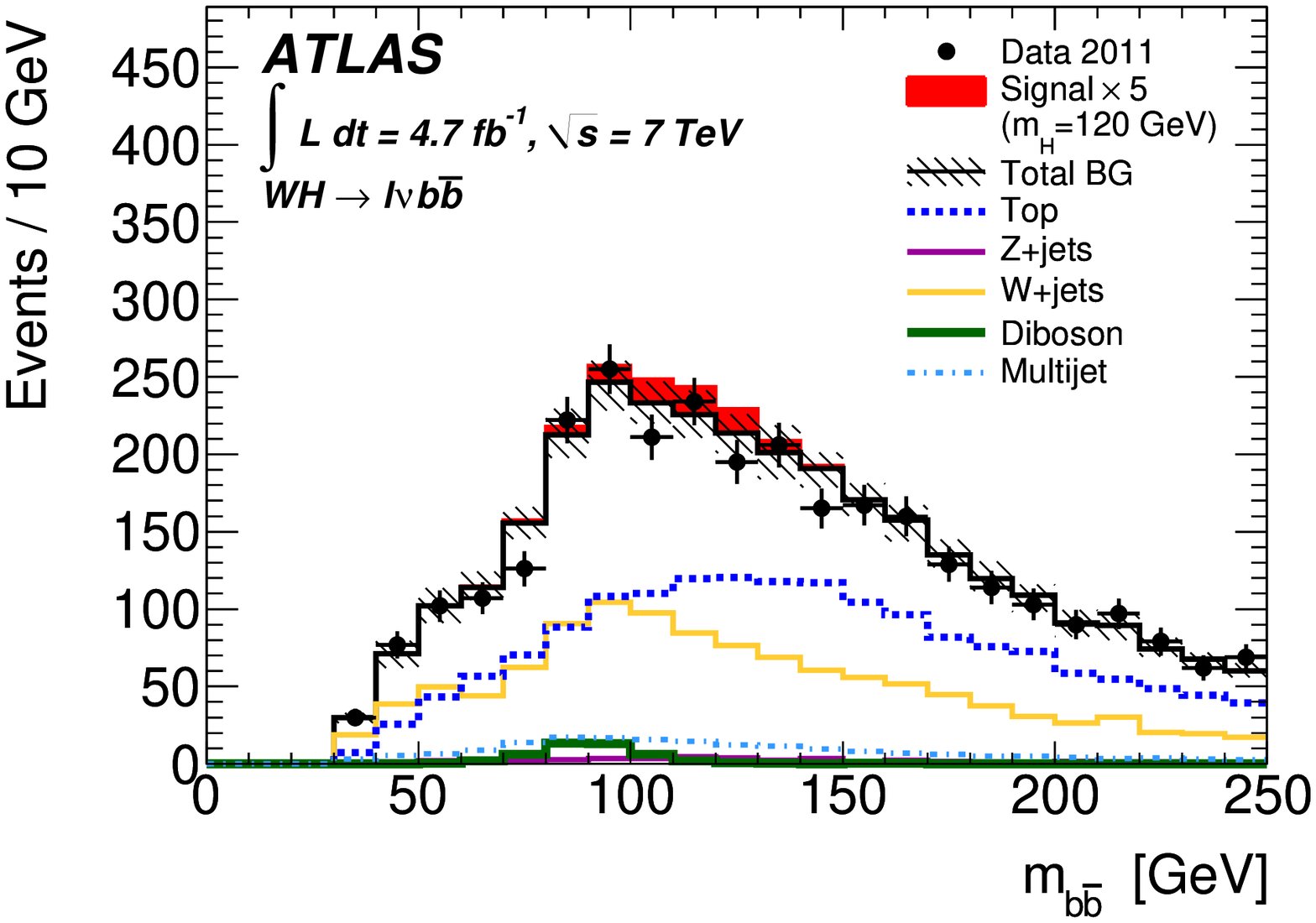,width=\figsize}}%
           \hspace{\columnsep}%
           \parbox[t]{\figsize}{\vskip 0.1pt \epsfig{file=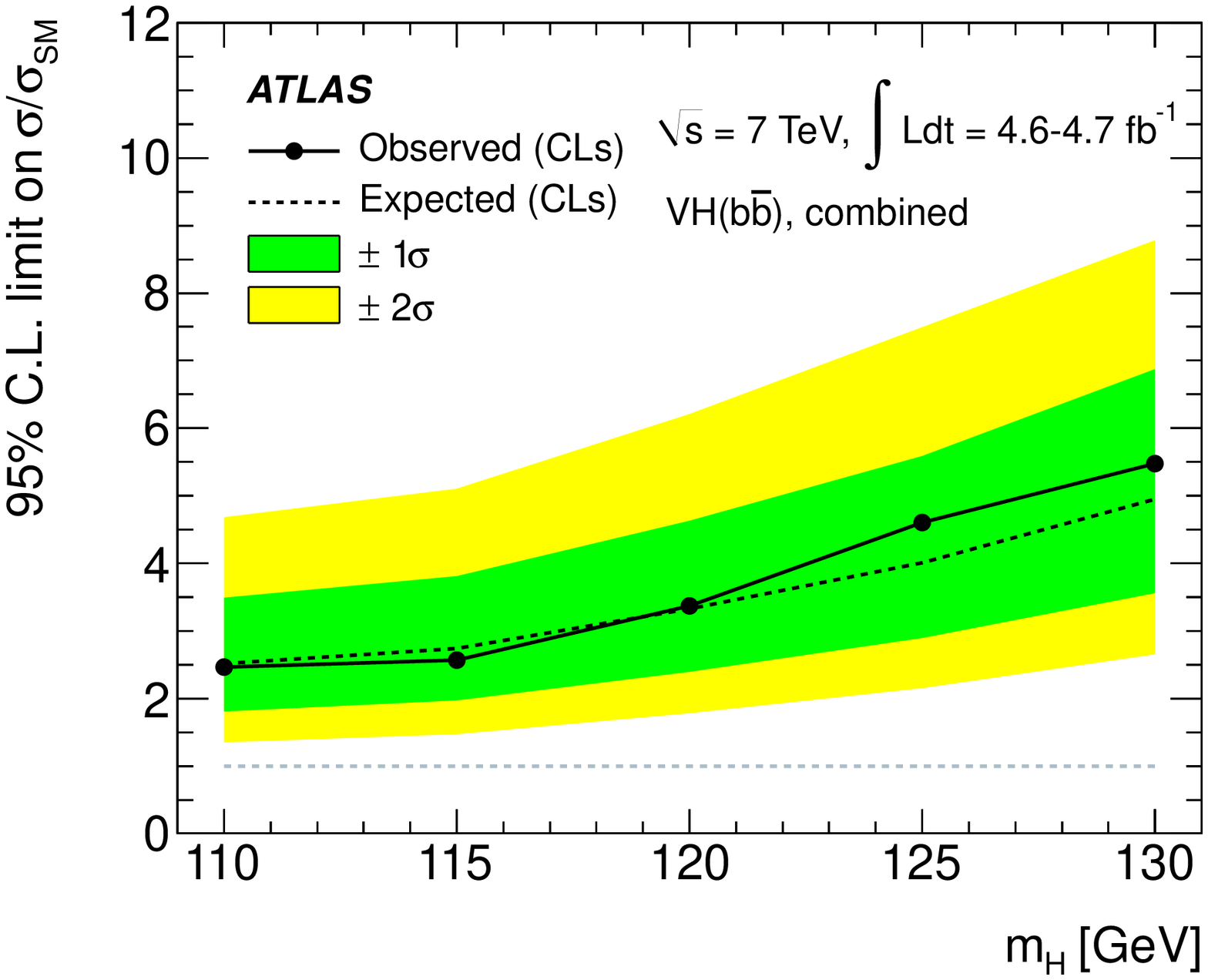,width=\figsize}}}}}}%
{{%
  \newdimen\figsize
  \setlength\figsize{\hsize}%
  \addtolength\figsize{-\columnsep}%
  \addtolength\figsize{-\columnsep}%
  \divide\figsize by 3
  \vbox{%
  \makebox{\parbox[t]{\figsize}{\vskip 0.1pt \epsfig{file=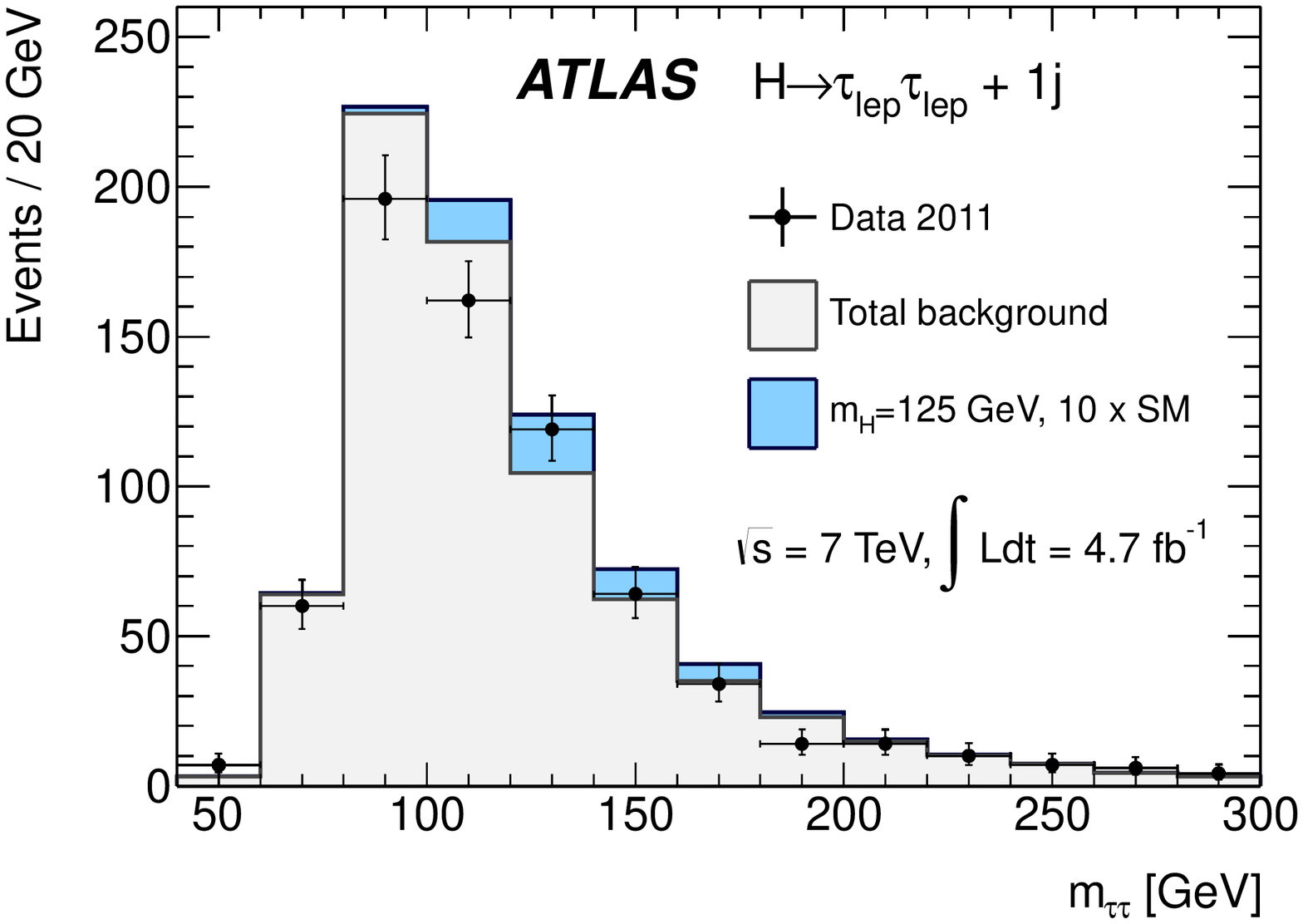,width=\figsize,height=3.8cm}}%
           \hspace{\columnsep}%
           \parbox[t]{\figsize}{\vskip 0.1pt \epsfig{file=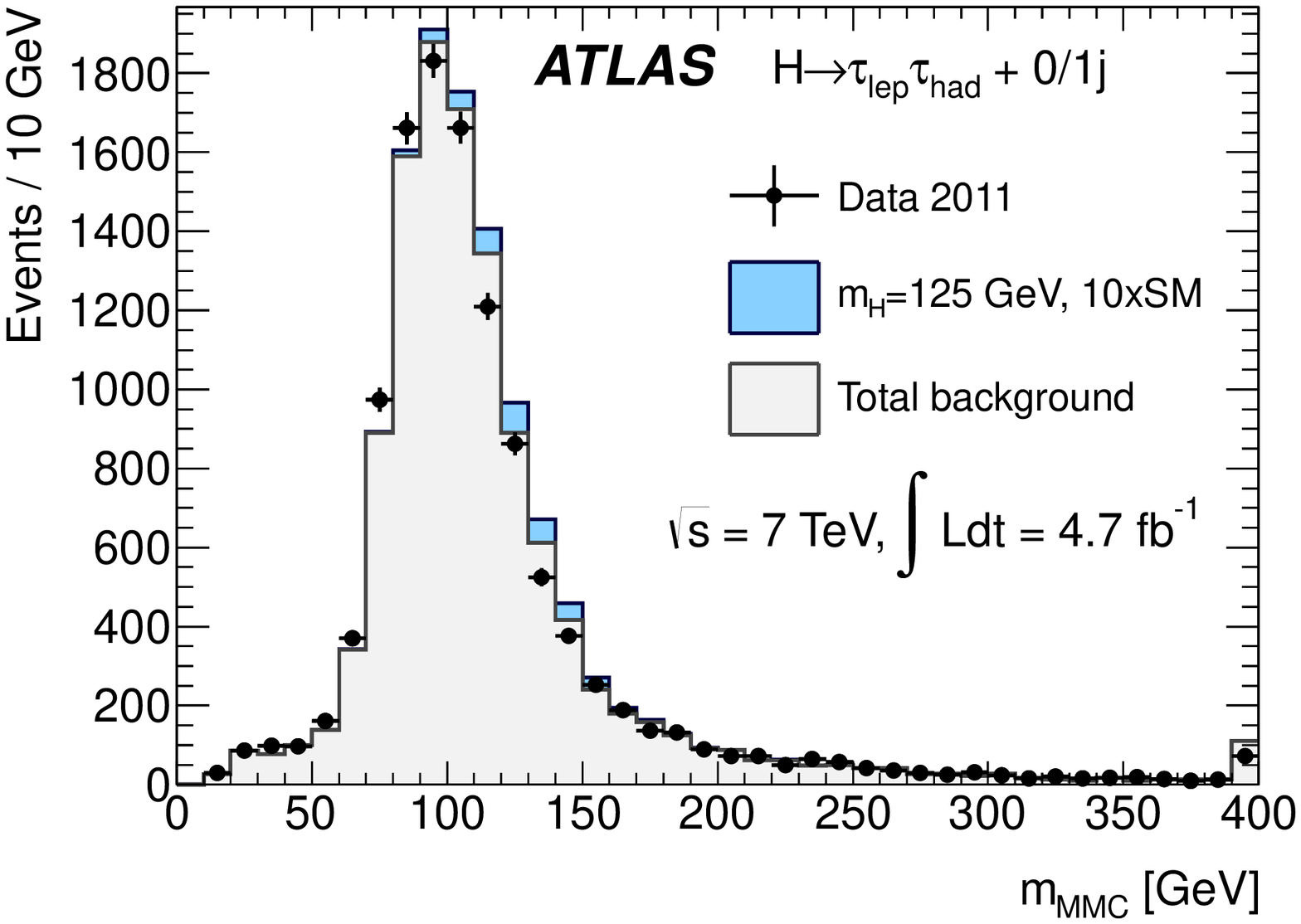,width=\figsize,height=3.8cm}}%
           \hspace{\columnsep}%
           \parbox[t]{\figsize}{\vskip 0.1pt \epsfig{file=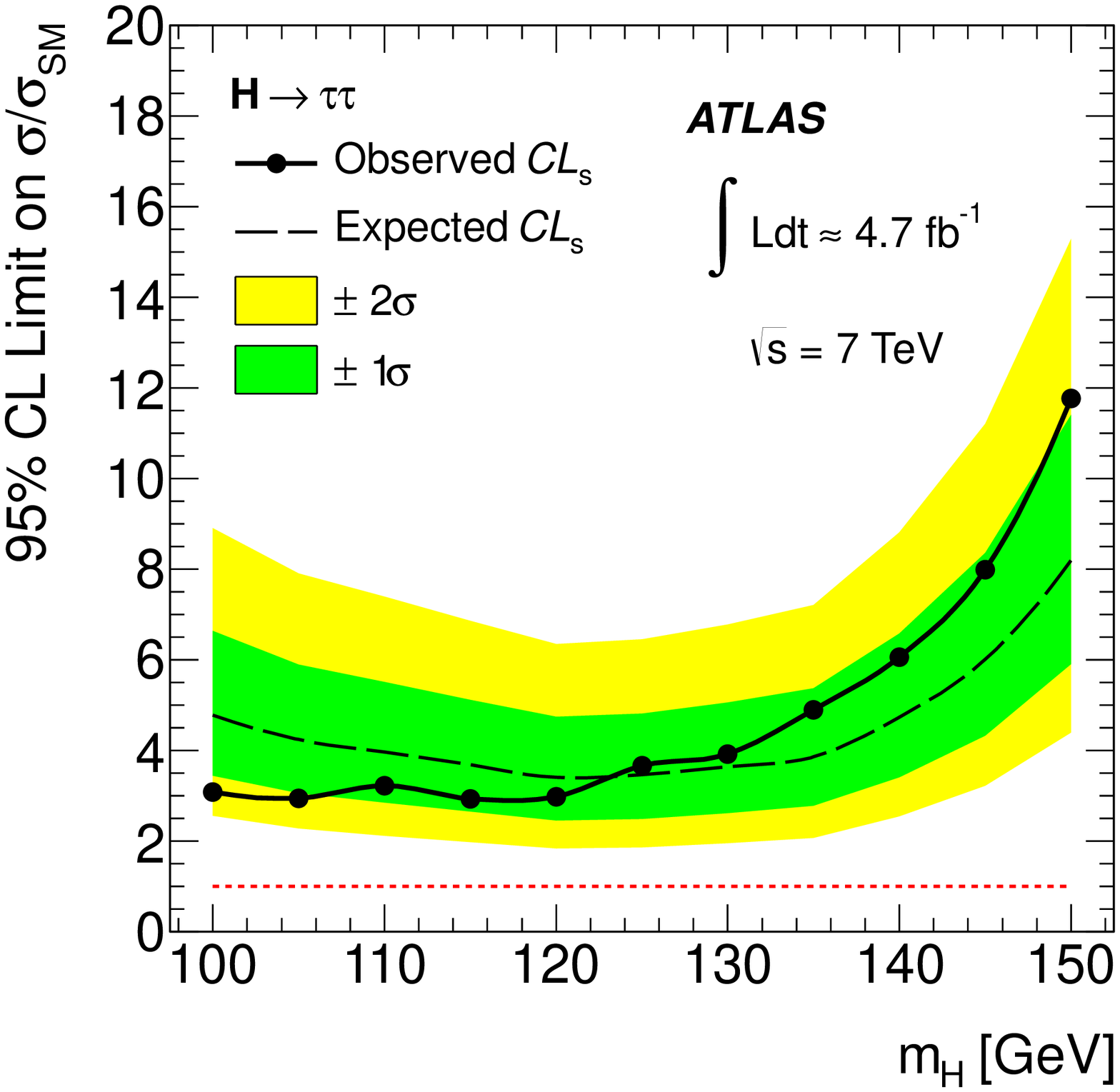,width=\figsize,height=3.8cm}}}}}}%
\caption{Selected results.  Left two columns: data compared with
  expected background for (Row 1) $m_{4\ell}$ from the 
  $H{\rightarrow} ZZ^{(*)}{\rightarrow} \ell\ell\ell'\ell'$ channel for $100{\ifmmode {\mathrm{\ Ge\kern -0.1em V}}\else\textrm{Ge\kern -0.1em V}\fi}<{\ensuremath{m_H}}<250{\ifmmode {\mathrm{\ Ge\kern -0.1em V}}\else\textrm{Ge\kern -0.1em V}\fi}$
  and for the complete range~\cite{ATLAS:2012ac}; (Row 2) $m_T$ from the
  $H{\rightarrow} WW^{(*)}{\rightarrow} \ell\nu\ell\nu$ channel with 0/1 jets~\cite{ATLAS:2012sc};
  (Row 3) $m_{bb}$ from the $ZH$ and $WH$ channels~\cite{ATLAS:2012zf};
  (Row 4) $m_{\tau\tau}$ from the $H{\rightarrow}{\ensuremath{\tau_{\mathrm{lep}}}}{\ensuremath{\tau_{\mathrm{lep}}}}+1j$ and
  $H{\rightarrow}{\ensuremath{\tau_{\mathrm{lep}}}}{\ensuremath{\tau_{\mathrm{had}}}}+0/1j$ channels~\cite{ATLAS:2012ur}.
  An example of the expected signal is also shown (scaled up for some
  channels to make it more visible).
  Right: Corresponding exclusion plots, showing the
  expected (dashed) and observed (solid) 95\% CL upper limits
  on SM Higgs boson production.  The dark (green) and light (yellow)
  bands indicate the expected limits with $\pm1\sigma$ and $\pm2\sigma$
  fluctuations, respectively.}
\label{fig:results1}
\end{center}
\end{figure}

The results are shown in Figure~\ref{fig:results1}, Row 1.  The ${\ensuremath{m_H}}$ regions
$(134$--$156){\ifmmode {\mathrm{\ Ge\kern -0.1em V}}\else\textrm{Ge\kern -0.1em V}\fi}$, $(182$--$233){\ifmmode {\mathrm{\ Ge\kern -0.1em V}}\else\textrm{Ge\kern -0.1em V}\fi}$, $(256$--$265){\ifmmode {\mathrm{\ Ge\kern -0.1em V}}\else\textrm{Ge\kern -0.1em V}\fi}$, and
$(268$--$415){\ifmmode {\mathrm{\ Ge\kern -0.1em V}}\else\textrm{Ge\kern -0.1em V}\fi}$ are excluded at 95\% CL.  Small excesses
are seen around $125{\ifmmode {\mathrm{\ Ge\kern -0.1em V}}\else\textrm{Ge\kern -0.1em V}\fi}$, $244{\ifmmode {\mathrm{\ Ge\kern -0.1em V}}\else\textrm{Ge\kern -0.1em V}\fi}$, and $500{\ifmmode {\mathrm{\ Ge\kern -0.1em V}}\else\textrm{Ge\kern -0.1em V}\fi}$ with
local significances of $2.1\sigma$, $2.2\sigma$, and $2.1\sigma$,
respectively.  The excess at $125{\ifmmode {\mathrm{\ Ge\kern -0.1em V}}\else\textrm{Ge\kern -0.1em V}\fi}$ corresponds to three
events that cluster within the ${\ensuremath{m_H}}$ mass resolution of $2\%$:
two $2e2\mu$ events at 123.6 and $124.3{\ifmmode {\mathrm{\ Ge\kern -0.1em V}}\else\textrm{Ge\kern -0.1em V}\fi}$, and one $4\mu$
event at $124.3{\ifmmode {\mathrm{\ Ge\kern -0.1em V}}\else\textrm{Ge\kern -0.1em V}\fi}$.

\subsection{$H{\rightarrow} WW^{(*)}{\rightarrow} \ell\nu\ell\nu$ channel}

This channel~\cite{ATLAS:2012sc} is also sensitive over a wide mass range.
The event selection requires two isolated, opposite-sign leptons
and a large missing transverse energy (${\ensuremath{E_{\mathrm{T}}^{\mathrm{miss}}}}$).  The analysis is divided 
into subsamples with 0, 1, and $\ge 2$ jets, as the background
composition is quite different for these cases:
for the 0-jet case, the background
is dominated by the $WW$ and $Z+{\mathrm{jets}}$ processes, while for the 
2-jet case, ${t\bar t}$ dominates.  As there are two neutrinos in the
final state, the Higgs~boson mass cannot be reconstructed; the
transverse mass $m_T$ is used instead.  Results are shown
in Figure~\ref{fig:results1}, Row 2.  No excess is seen, and the ${\ensuremath{m_H}}$ region
$(133$--$261){\ifmmode {\mathrm{\ Ge\kern -0.1em V}}\else\textrm{Ge\kern -0.1em V}\fi}$ is excluded at 95\% CL.

\subsection{Other low-${\ensuremath{m_H}}$ channels}

Two additional channels are sensitive to a low-mass
Higgs~boson.  In the $V(H{\rightarrow} b\bar b)$ channel~\cite{ATLAS:2012zf},
one searches for
a Higgs~boson produced in association with a vector gauge boson,
with the Higgs~boson decaying into a $b\bar b$ pair.  The event selection
for this channel requires a gauge boson decay to leptons/neutrinos (one of
$W{\rightarrow}\ell\nu$, $Z{\rightarrow}\ell\ell$, or $Z{\rightarrow}\nu\nu$) and exactly
two $b$-tagged jets.  The analysis is subdivided in bins
of ${\ensuremath{p_{\mathrm{T}}}}(V)$.  The $H{\rightarrow}\tau\tau$ analysis~\cite{ATLAS:2012ur} is divided into
$2\ell4\nu$, $\ell{\ensuremath{\tau_{\mathrm{had}}}}3\nu$, and $2{\ensuremath{\tau_{\mathrm{had}}}}2\nu$ subchannels;
these are then further subdivided according to the number of extra
jets in the event.  Since there are multiple neutrinos in the final state,
the invariant mass $m_{\tau\tau}$
is estimated either using the collinear approximation (assuming
the neutrino to be collinear with the visible decay products)
or the ``missing mass calculator'' technique (incorporating
the probability distribution for the opening angle in $\tau$ decays).

Selected results are shown in Figure~\ref{fig:results1}, Rows 3 and 4.
Exclusion limits for these channels range from about 2.5 to 12 times
the SM Higgs~boson production cross section over the ${\ensuremath{m_H}}$ range
$(100$--$150){\ifmmode {\mathrm{\ Ge\kern -0.1em V}}\else\textrm{Ge\kern -0.1em V}\fi}$.

\subsection{Other high-${\ensuremath{m_H}}$ (diboson) channels}

The remaining diboson channels are sensitive to higher Higgs~boson
masses.  For the $H{\rightarrow} ZZ{\rightarrow} \ell\ell\nu\nu$ channel~\cite{ATLAS:2012va},
events with
a $Z{\rightarrow}\ell\ell$ decay and large ${\ensuremath{E_{\mathrm{T}}^{\mathrm{miss}}}}$ are selected.
Different selection requirements are made for ${\ensuremath{m_H}}$ above or below $280{\ifmmode {\mathrm{\ Ge\kern -0.1em V}}\else\textrm{Ge\kern -0.1em V}\fi}$.
For $H{\rightarrow} ZZ{\rightarrow} \ell\ell q q$~\cite{ATLAS:2012vw},
events are selected with one $Z{\rightarrow}\ell\ell$
decay, one $Z{\rightarrow} jj$ decay, and small ${\ensuremath{E_{\mathrm{T}}^{\mathrm{miss}}}}$.
As the fraction of heavy-flavour in the final state is quite different
for signal and background, this channel is divided into ``tagged''
(with two identified $b$-jets) and ``untagged'' subchannels.
The principal backgrounds for these two channels are $Z+{\mathrm{jets}}$, ${t\bar t}$,
and $ZZ$.  For the $H{\rightarrow} WW{\rightarrow}\ell\nu qq$ channel~\cite{ATLAS:2012vz},
events are selected
with an isolated lepton, large ${\ensuremath{E_{\mathrm{T}}^{\mathrm{miss}}}}$, and a $W{\rightarrow} jj$ decay.  This
analysis is divided into subchannels according to the number 
of additional jets in the event.

\begin{figure}[htb]
\begin{center}
{{%
  \newdimen\figsize
  \setlength\figsize{\hsize}%
  \addtolength\figsize{-\columnsep}%
  \addtolength\figsize{-\columnsep}%
  \divide\figsize by 3
  \vbox{%
  \makebox{\parbox[t]{\figsize}{\vskip 0.1pt \epsfig{file=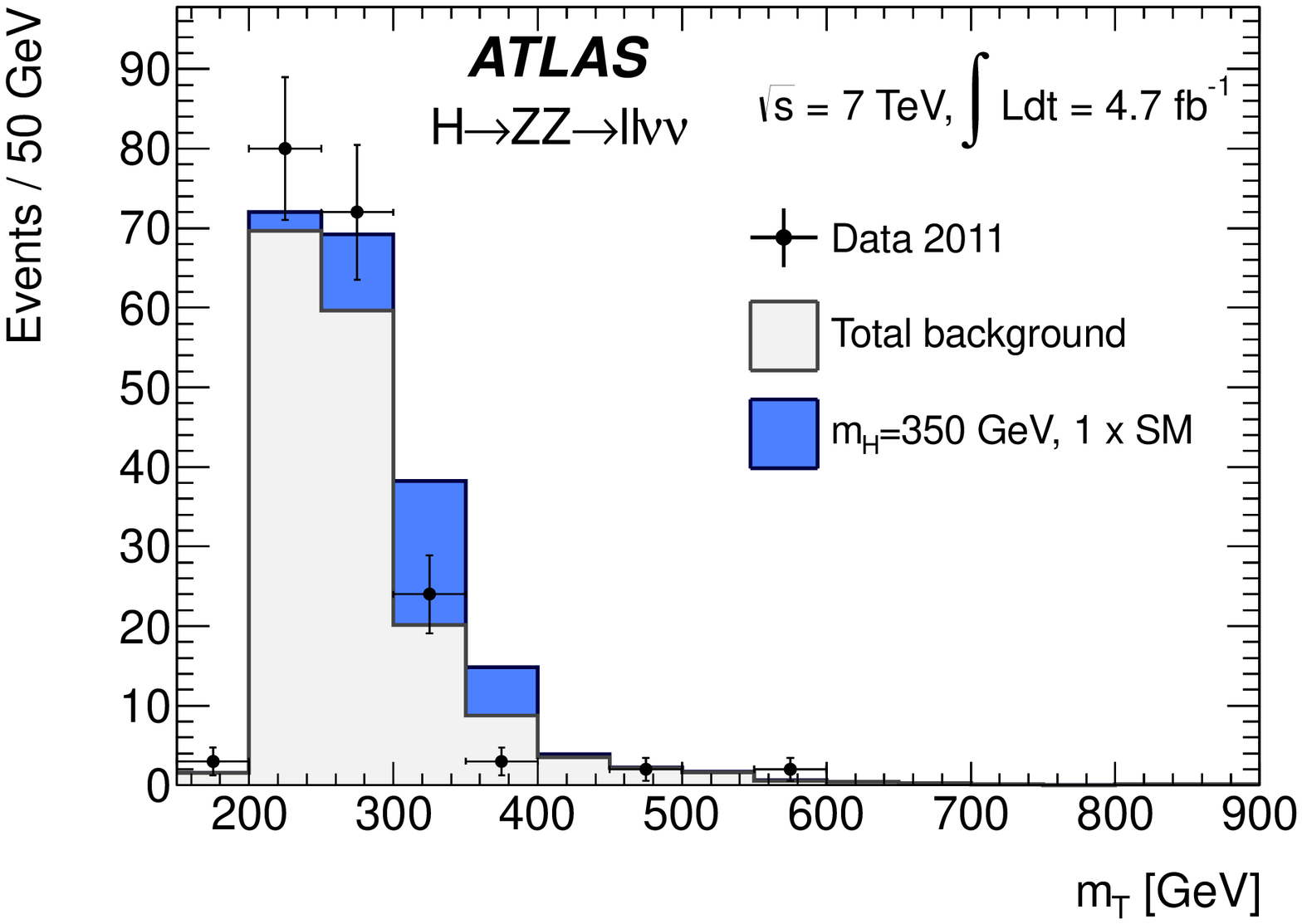,width=\figsize,height=3.0cm}}%
           \hspace{\columnsep}%
           \parbox[t]{\figsize}{\vskip 0.1pt \epsfig{file=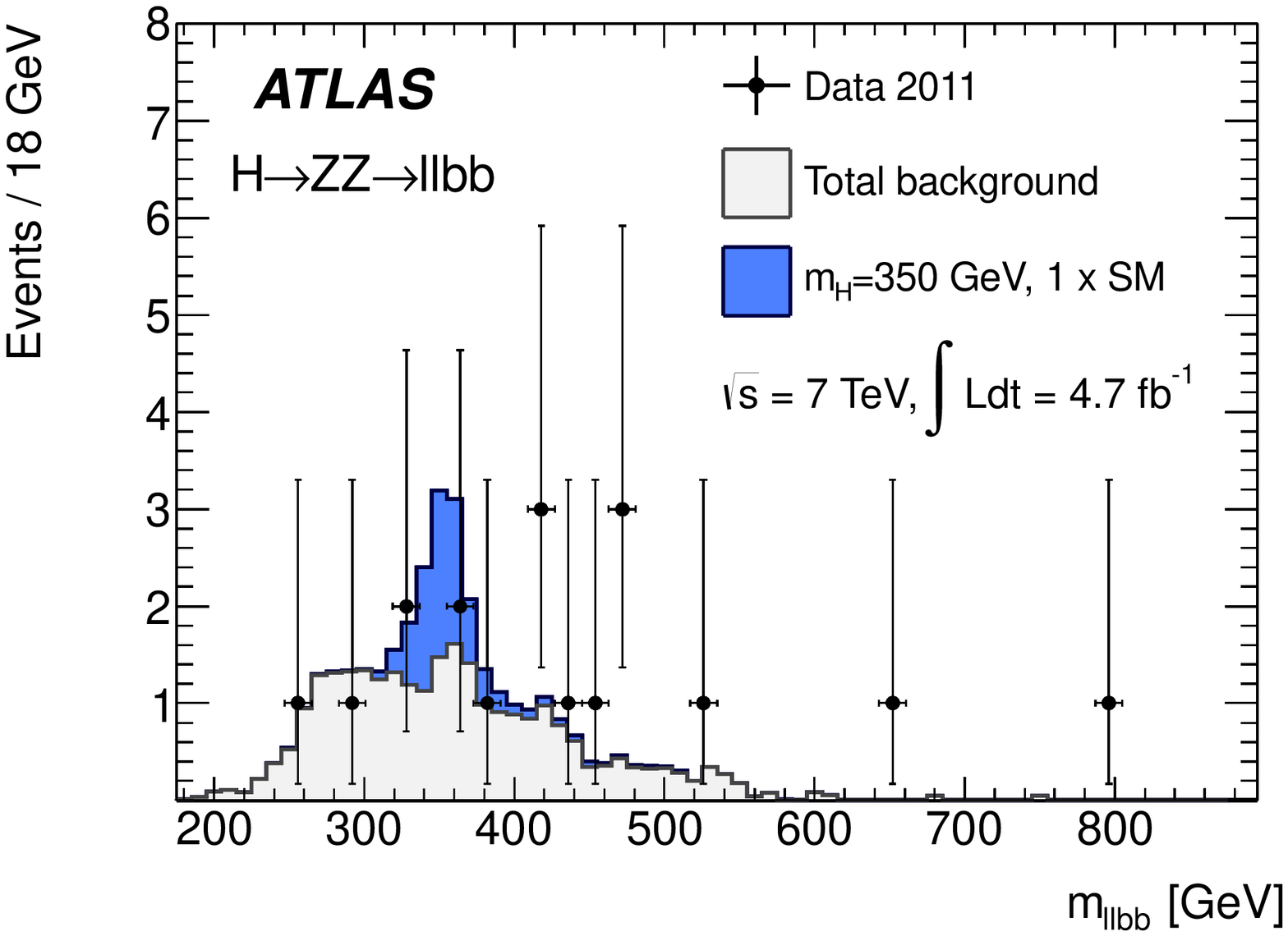,width=\figsize,height=3.0cm}}%
           \hspace{\columnsep}%
           \parbox[t]{\figsize}{\vskip 0.1pt \epsfig{file=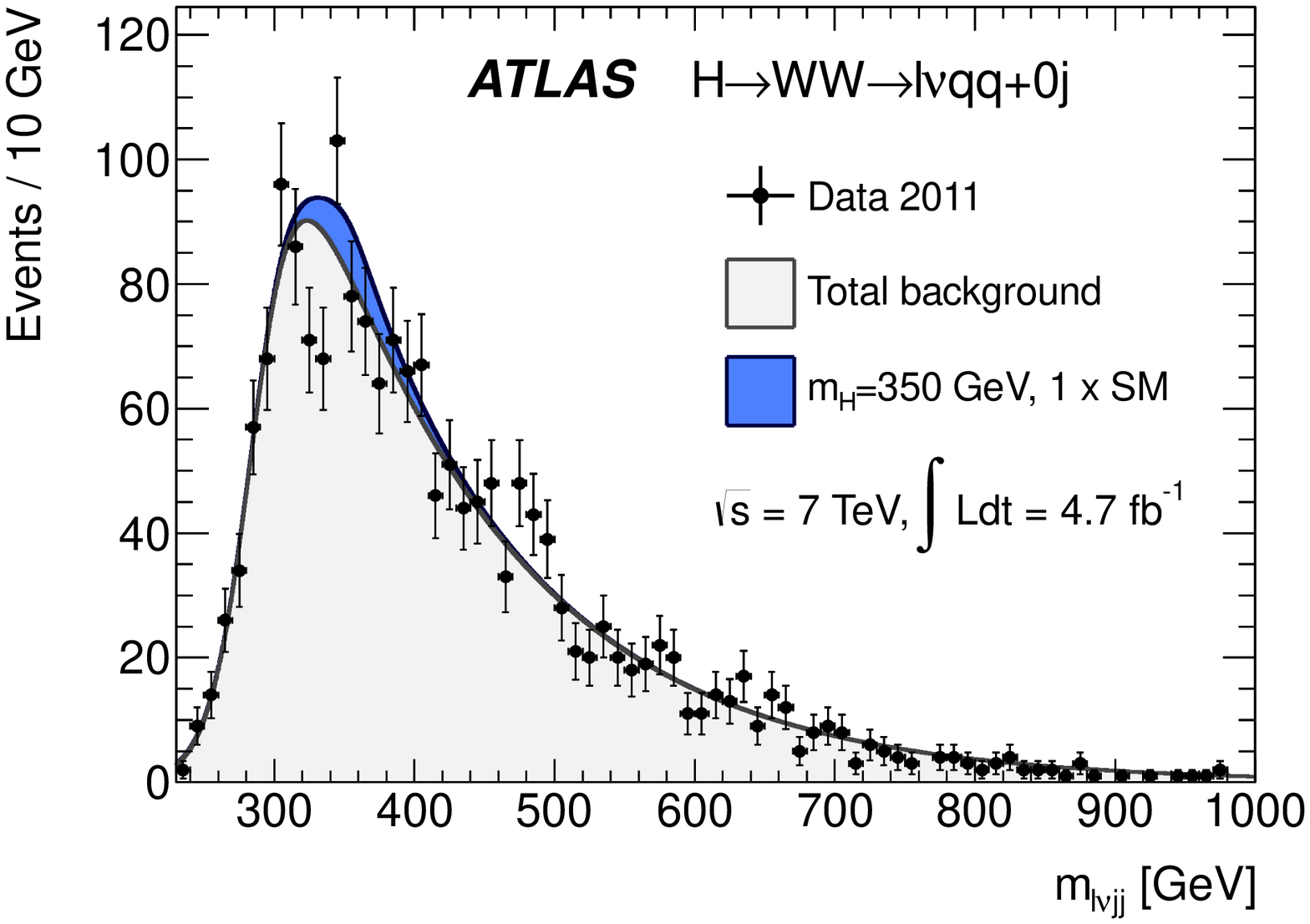,width=\figsize,height=3.0cm}}}}}}%
{{%
  \newdimen\figsize
  \setlength\figsize{\hsize}%
  \addtolength\figsize{-\columnsep}%
  \addtolength\figsize{-\columnsep}%
  \divide\figsize by 3
  \vbox{%
  \makebox{\parbox[t]{\figsize}{\vskip 0.1pt \epsfig{file=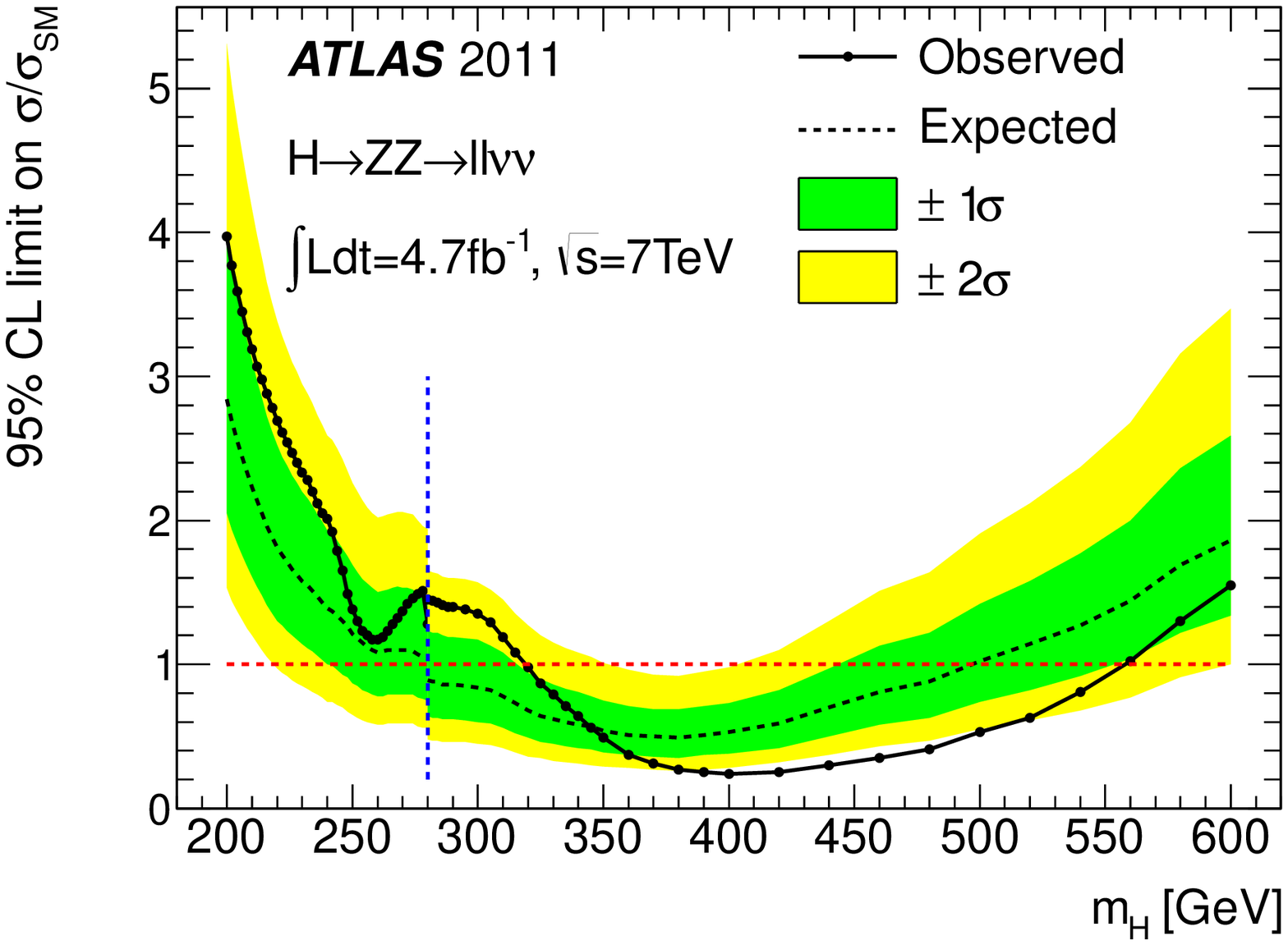,width=\figsize,height=3.0cm}}%
           \hspace{\columnsep}%
           \parbox[t]{\figsize}{\vskip 0.1pt \epsfig{file=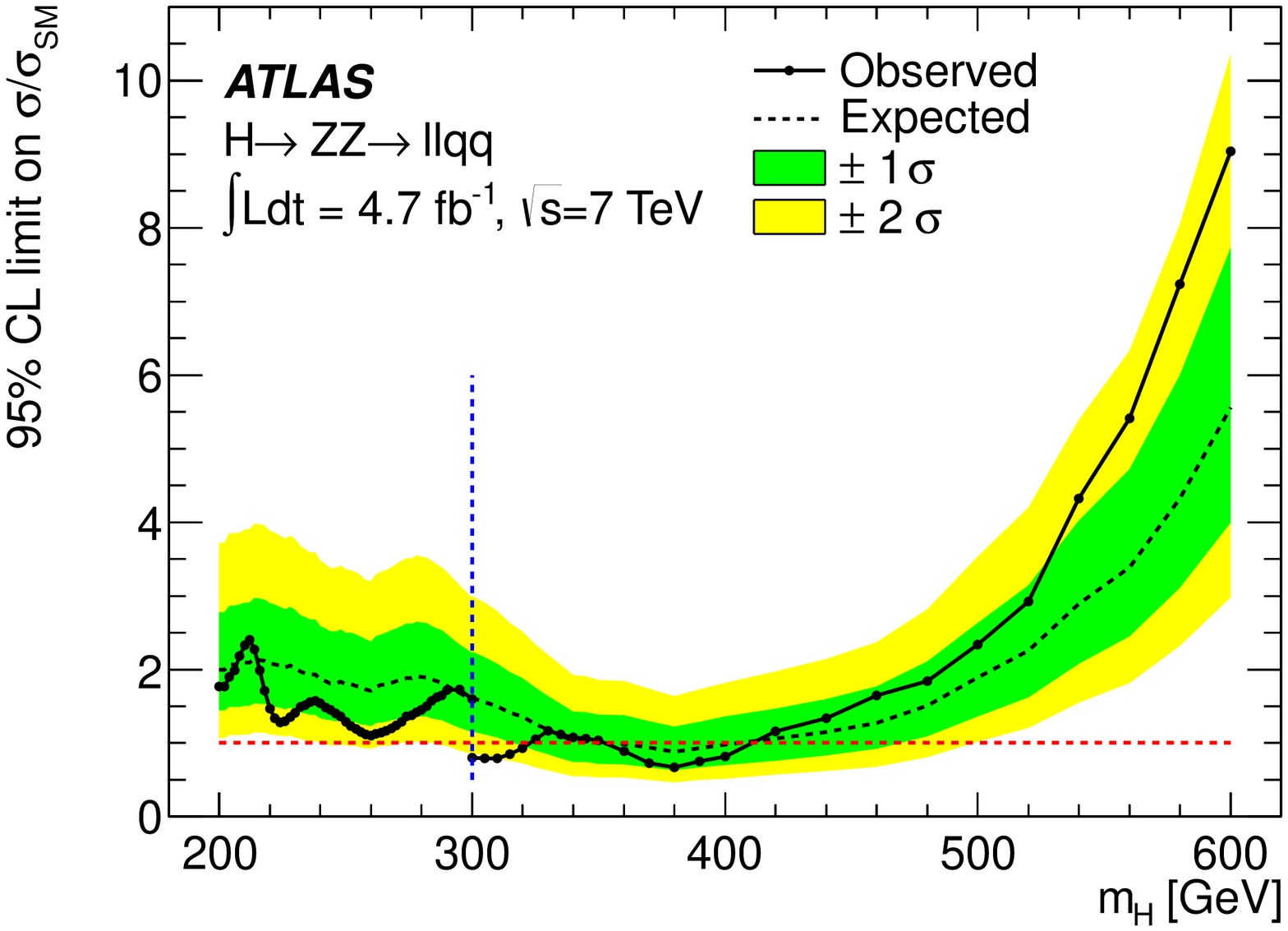,width=\figsize,height=3.0cm}}%
           \hspace{\columnsep}%
           \parbox[t]{\figsize}{\vskip 0.1pt \epsfig{file=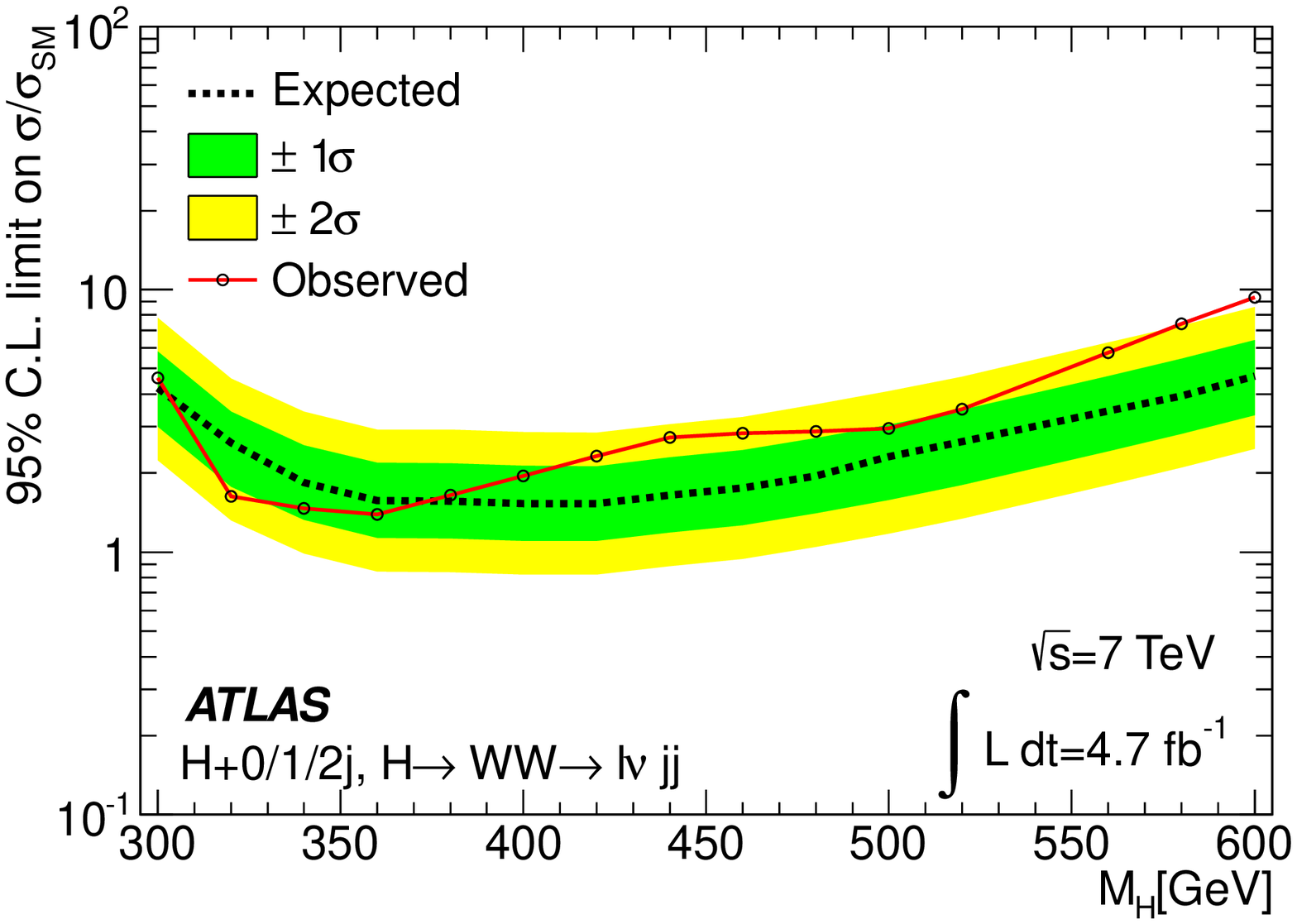,width=\figsize,height=3.0cm}}}}}}%
\caption{Selected results from the remaining diboson channels.
  Top: data and expected background for
  $H{\rightarrow} ZZ{\rightarrow} \ell\ell\nu\nu$, showing $m_T$~\cite{ATLAS:2012va} (left);
  $H{\rightarrow} ZZ{\rightarrow} \ell\ell q q$, showing $m_{\ell\ell bb}$ for the high-${\ensuremath{m_H}}$,
  tagged subchannel~\cite{ATLAS:2012vw} (middle); and 
  $H{\rightarrow} WW{\rightarrow}\ell\nu qq$, showing $m_{\ell\nu jj}$ for the 0-jet
  subchannel~\cite{ATLAS:2012vz} (right).
  The expectations for a Higgs boson of ${\ensuremath{m_H}}=350{\ifmmode {\mathrm{\ Ge\kern -0.1em V}}\else\textrm{Ge\kern -0.1em V}\fi}$ are also shown.
  Bottom: Corresponding exclusion plots (see Figure~\ref{fig:results1}).}
\label{fig:otherzz}
\end{center}
\end{figure}

Results are shown in Figure~\ref{fig:otherzz}.
The $ZZ{\rightarrow} \ell\ell\nu\nu$ channel excludes at 95\% CL the ${\ensuremath{m_H}}$ range
$(320$--$560){\ifmmode {\mathrm{\ Ge\kern -0.1em V}}\else\textrm{Ge\kern -0.1em V}\fi}$, while the $ZZ{\rightarrow}\ell\ell qq$ channel excludes
the ranges $(300$--$322){\ifmmode {\mathrm{\ Ge\kern -0.1em V}}\else\textrm{Ge\kern -0.1em V}\fi}$ and $(353$--$410){\ifmmode {\mathrm{\ Ge\kern -0.1em V}}\else\textrm{Ge\kern -0.1em V}\fi}$.  Exclusions
from the $WW{\rightarrow}\ell\nu qq$ channel are 2--10 times the SM
cross section.

\section{Combination of all channels}

\begin{figure}[tb]
\begin{center}
{{%
  \newdimen\figsize
  \setlength\figsize{\hsize}%
  \addtolength\figsize{-\columnsep}%
  \addtolength\figsize{-\columnsep}%
  \divide\figsize by 3
  \vbox{%
  \makebox{\parbox[t]{\figsize}{\vskip 0.1pt \epsfig{file=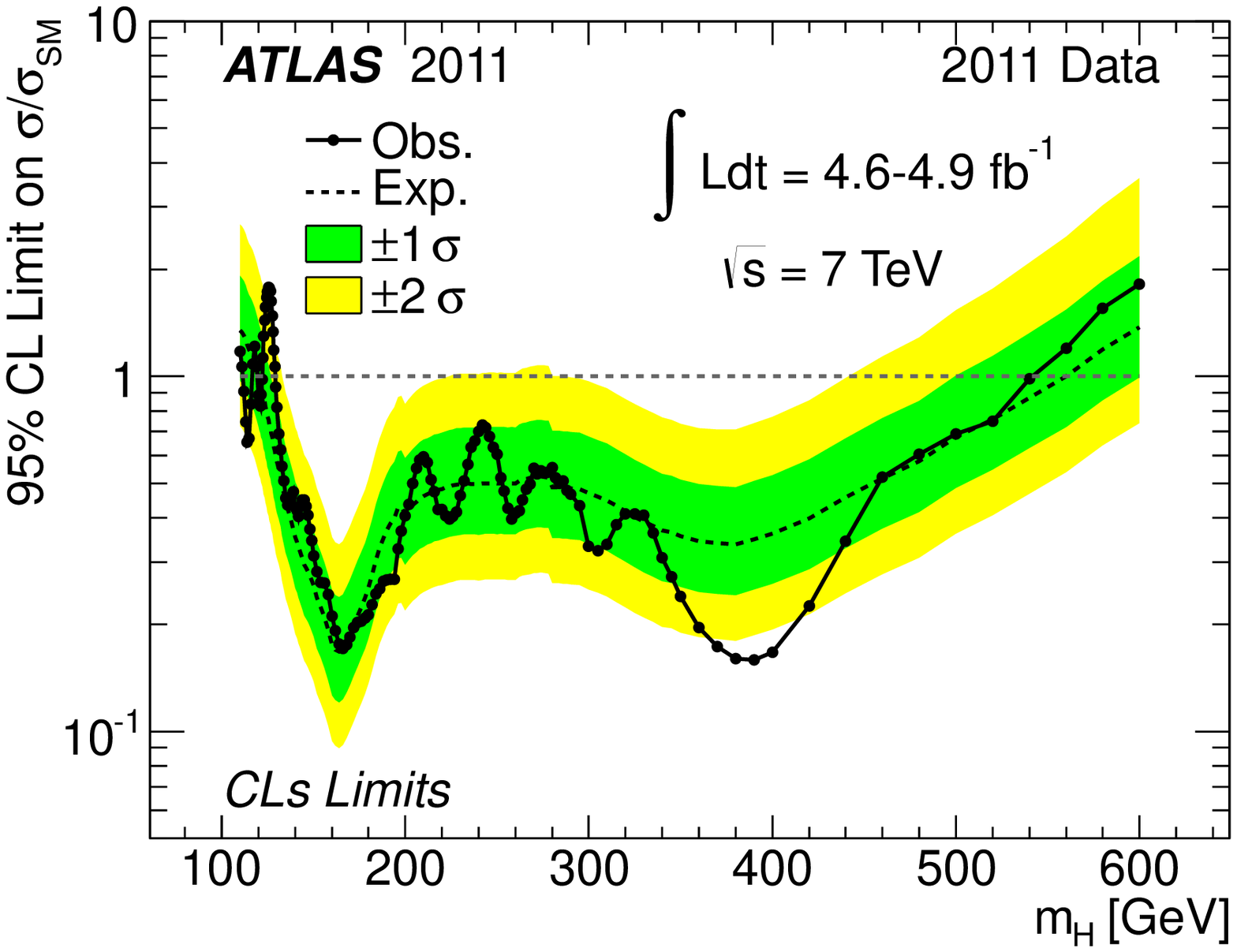,width=\figsize,height=3.6cm}}%
           \hspace{\columnsep}%
           \parbox[t]{\figsize}{\vskip 0.1pt \epsfig{file=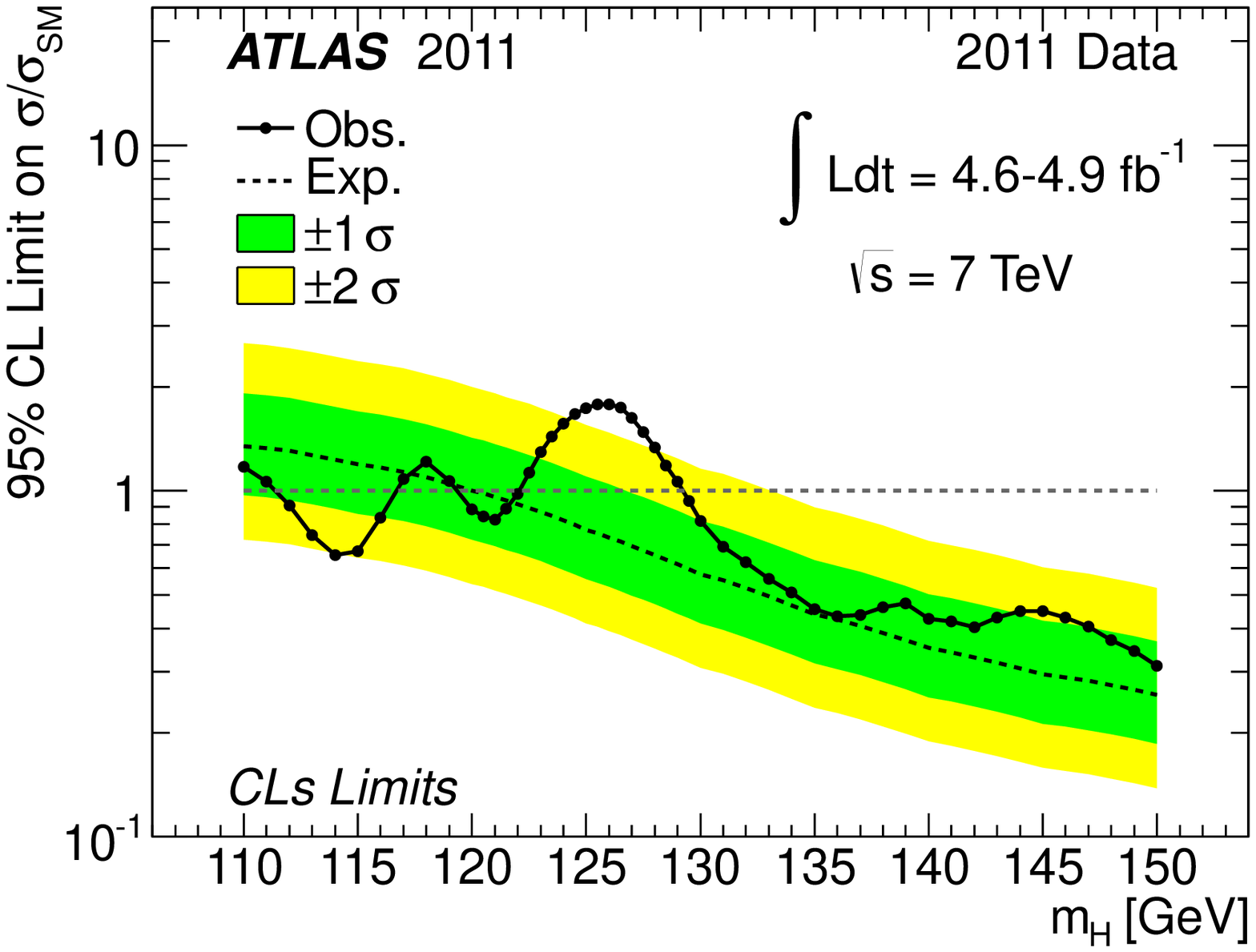,width=\figsize,height=3.6cm}}%
           \hspace{\columnsep}%
           \parbox[t]{\figsize}{\vskip 0.1pt \epsfig{file=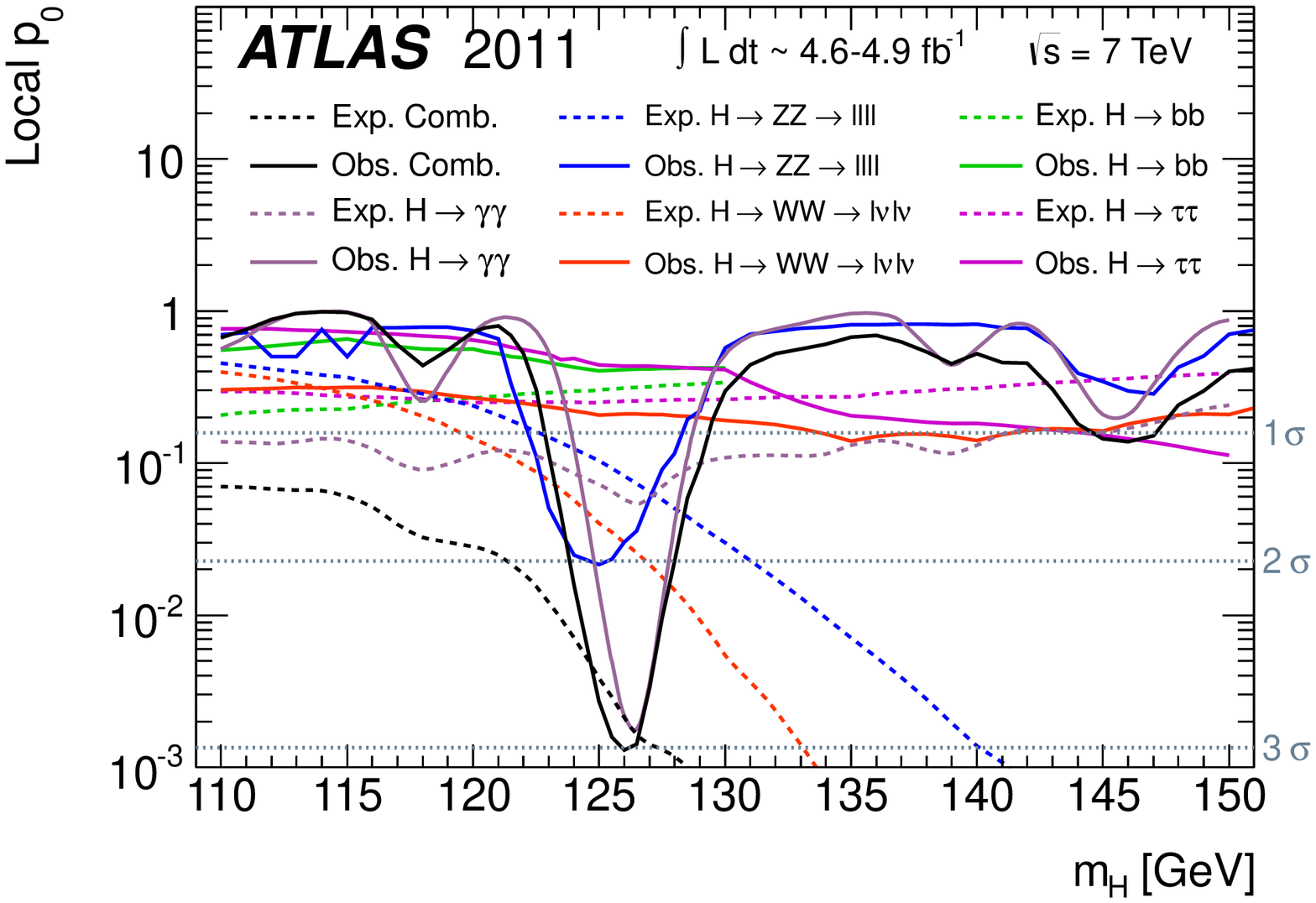,width=\figsize,height=3.6cm}}}}}}%
\caption{Left, middle: Combined exclusion limits for the full mass
  range and for ${\ensuremath{m_H}}<150{\ifmmode {\mathrm{\ Ge\kern -0.1em V}}\else\textrm{Ge\kern -0.1em V}\fi}$~\cite{ATLAS:2012an}
  (see Figure~\ref{fig:results1}).
  Right: Local probability $p_0$ for the background
  to fluctuate to the observed number of events or higher, by channel,
  for ${\ensuremath{m_H}}<150{\ifmmode {\mathrm{\ Ge\kern -0.1em V}}\else\textrm{Ge\kern -0.1em V}\fi}$.  Dashed
  lines show the expected median local $p_0$ for the signal hypotheses
  at ${\ensuremath{m_H}}$.
}
\label{fig:comb}
\end{center}
\end{figure}

For the final results, all these channels are combined~\cite{ATLAS:2012an}.
Systematic uncertainties
are taken to be either 100\% correlated or uncorrelated between channels.
The results are shown in Figure~\ref{fig:comb}.  The ${\ensuremath{m_H}}$ ranges
$(111.4$--$116.6){\ifmmode {\mathrm{\ Ge\kern -0.1em V}}\else\textrm{Ge\kern -0.1em V}\fi}$, $(119.4$--$122.1){\ifmmode {\mathrm{\ Ge\kern -0.1em V}}\else\textrm{Ge\kern -0.1em V}\fi}$, and $(129.2$--$541){\ifmmode {\mathrm{\ Ge\kern -0.1em V}}\else\textrm{Ge\kern -0.1em V}\fi}$ are 
excluded at 95\% CL, and the ${\ensuremath{m_H}}$ range $(130.7$--$506){\ifmmode {\mathrm{\ Ge\kern -0.1em V}}\else\textrm{Ge\kern -0.1em V}\fi}$ is excluded
at 99\%.  An excess is seen around ${\ensuremath{m_H}}=126{\ifmmode {\mathrm{\ Ge\kern -0.1em V}}\else\textrm{Ge\kern -0.1em V}\fi}$, with a local significance
of $3.0\sigma$.  The global significance is approximately 15\% over
the full range $(100$--$600){\ifmmode {\mathrm{\ Ge\kern -0.1em V}}\else\textrm{Ge\kern -0.1em V}\fi}$ and (5--7)\% over the range
$(110$--$146){\ifmmode {\mathrm{\ Ge\kern -0.1em V}}\else\textrm{Ge\kern -0.1em V}\fi}$,
corresponding to the range at low-${\ensuremath{m_H}}$ not excluded by the previous
LHC SM Higgs boson combined search at 99\% CL~\cite{atlascms}.
If the excess is interpreted as a SM Higgs boson, the corresponding
cross section ratio at ${\ensuremath{m_H}}=126{\ifmmode {\mathrm{\ Ge\kern -0.1em V}}\else\textrm{Ge\kern -0.1em V}\fi}$ is
$\mu = \sigma_{\mathrm{obs}} / \sigma_{\mathrm{SM}} = 1.1\pm0.4$,
consistent with the SM value of 1.0.  The excess is observed only
in the $H{\rightarrow}\gamma\gamma$ and $H{\rightarrow}\ell\ell\ell'\ell'$ channels;
however, none of the other channels are inconsistent with a SM
Higgs~boson at this mass.

\section{Summary}

From the 2011 data, ATLAS excludes at 95\% CL the ${\ensuremath{m_H}}$ range
$(111.4$--$541){\ifmmode {\mathrm{\ Ge\kern -0.1em V}}\else\textrm{Ge\kern -0.1em V}\fi}$, except for the regions $(116.6$--$119.4){\ifmmode {\mathrm{\ Ge\kern -0.1em V}}\else\textrm{Ge\kern -0.1em V}\fi}$
and $(122.1$--$129.2){\ifmmode {\mathrm{\ Ge\kern -0.1em V}}\else\textrm{Ge\kern -0.1em V}\fi}$.  An excess is seen in the $H{\rightarrow}\gamma\gamma$
and $H{\rightarrow}\ell\ell\ell'\ell'$ channels at ${\ensuremath{m_H}}\approx126{\ifmmode {\mathrm{\ Ge\kern -0.1em V}}\else\textrm{Ge\kern -0.1em V}\fi}$
which is consistent with a SM Higgs~boson.  The chance of this
being due to a background fluctuation is 15\% over the full mass
range, or (5--7)\% over $(110$--$146){\ifmmode {\mathrm{\ Ge\kern -0.1em V}}\else\textrm{Ge\kern -0.1em V}\fi}$.

\bigskip
This work is supported in part by the U.S.~Department of Energy under
contract DE-AC02-98CH10886 with Brookhaven National Laboratory.

\end{document}